\documentclass[12pt,preprint]{emulateapj}
\usepackage{color}

\def\gsim{\mathrel{\rlap{\lower 4pt \hbox{\hskip 1pt $\sim$}}\raise 1pt \hbox {$>$}}} 
\def\lsim{\mathrel{\rlap{\lower 4pt \hbox{\hskip 1pt $\sim$}}\raise 1pt \hbox {$<$}}}

\shorttitle{Type Iax Supernova 2012Z} \shortauthors{Yamanaka et al.}

\begin{document}

\title{OISTER Optical and Near-Infrared Observations of Type Iax Supernova 2012Z}

\author{Masayuki \textsc{Yamanaka}\altaffilmark{1,2,3,4},
        Keiichi \textsc{Maeda}\altaffilmark{5,6}, 
        Koji S. \textsc{Kawabata}\altaffilmark{3},
        Masaomi \textsc{Tanaka}\altaffilmark{7}, \\
        Nozomu \textsc{Tominaga}\altaffilmark{1,6}, 
        Hiroshi \textsc{Akitaya}\altaffilmark{3}, 
        Takahiro \textsc{Nagayama}\altaffilmark{8,9}, 
        Daisuke \textsc{Kuroda}\altaffilmark{10}, \\
        Jun \textsc{Takahashi}\altaffilmark{11}, 
        Yoshihiko \textsc{Saito}\altaffilmark{12},
	Kenshi \textsc{Yanagisawa}\altaffilmark{10},
	Akihiko \textsc{Fukui}\altaffilmark{10}, \\
        Ryo \textsc{Miyanoshita}\altaffilmark{9},
        Makoto \textsc{Watanabe}\altaffilmark{13},
        Akira \textsc{Arai}\altaffilmark{11,14},       
        Mizuki \textsc{Isogai}\altaffilmark{7,14},\\
        Takashi \textsc{Hattori}\altaffilmark{15},
	Hidekazu \textsc{Hanayama}\altaffilmark{16},
        Ryosuke \textsc{Itoh}\altaffilmark{4}, 
        Takahiro \textsc{Ui}\altaffilmark{4},\\
        Katsutoshi \textsc{Takaki}\altaffilmark{4},
        Issei \textsc{Ueno}\altaffilmark{4},
        Michitoshi \textsc{Yoshida}\altaffilmark{3},
        Gamal B. \textsc{Ali}\altaffilmark{17},\\
        Ahmed \textsc{Essam}\altaffilmark{17},
        Akihito \textsc{Ozaki}\altaffilmark{13},
        Hikaru \textsc{Nakao}\altaffilmark{13},
        Ko \textsc{Hamamoto}\altaffilmark{13},\\
	Daisaku \textsc{Nogami}\altaffilmark{2,5},
	Tomoki \textsc{Morokuma}\altaffilmark{18},
	Yumiko \textsc{Oasa}\altaffilmark{19},\\
        Hideyuki \textsc{Izumiura}\altaffilmark{10}, and
        Kazuhiro \textsc{Sekiguchi}\altaffilmark{7}.
}

\altaffiltext{1}{Department of Physics, Faculty of Science and Engineering, Konan University, Okamoto, Kobe, Hyogo 658-8501,
Japan; yamanaka@center.konan-u.ac.jp} 
\altaffiltext{2}{Kwasan Observatory, Kyoto University, 
17-1 Kitakazan-ohmine-cho, Yamashina-ku, Kyoto, 607-8471} 
\altaffiltext{3}{Hiroshima Astrophysical Science Center, Hiroshima University, Higashi-Hiroshima, Hiroshima 739-8526, Japan}
\altaffiltext{4}{Department of Physical Science, Hiroshima University, 1-3-1 Kagamiyama, Higashi-Hiroshima 739-8526, Japan} 
\altaffiltext{5}{Department of Astronomy, Graduate School of Science, Kyoto University, Sakyo-ku, Kyoto 606-8502, Japan}
\altaffiltext{6}{Institute for the Physics and Mathematics of the Universe (WPI), University of Tokyo, Kashiwa, Japan}
\altaffiltext{7}{National Astronomical Observatory of Japan, 2-21-1 Osawa, Mitaka, 
Tokyo 181-8588, Japan}
\altaffiltext{8}{Department of Astrophysics, Nagoya University, Chikusa-ku, Nagoya 464-8602, Japan}
\altaffiltext{9}{Graduate School of Science and Engineering, Kagoshima University, 1-21-35 Korimoto, Kagoshima 890-0065, Japan}
\altaffiltext{10}{Okayama Astrophysical Observatory, National Astronomical Observatory of Japan,  3037-5 Honjo, Kamogata, Asakuchi, Okayama 719-0232, Japan}
\altaffiltext{11}{Nishi-Harima Astronomical Observatory, Center for Astronomy, University of Hyogo, 407-2 Nishigaichi, Sayo-cho, Sayo, Hyogo 679-5313, Japan}
\altaffiltext{12}{Department of Physics, Tokyo Institute of Technology, 2-12-1 Ookayama, Meguro-ku, Tokyo 152-8551, Japan}
\altaffiltext{13}{Department of Cosmosciences, Hokkaido University, Kita 10, Nishi 8, Kita-ku, Sapporo, Hokkaido 060-0810, Japan}
\altaffiltext{14}{Koyama Astronomical Observatory, Kyoto Sangyo University, Motoyama, Kamigamo, Kita-Ku, Kyoto-City 603-8555, Japan}
\altaffiltext{15}{Subaru Telescope, 650 North Aʼohoku Place, Hilo, HI 96720, USA}
\altaffiltext{16}{Ishigakijima Astronomical Observatory, National Astronomical Observatory of Japan, 1024-1 Arakawa, Ishigaki, Okinawa 907-0024, Japan}
\altaffiltext{17}{National Research Institute of Astronomy and Geophysics, Cairo 11722, Egypt}
\altaffiltext{18}{Institute of Astronomy, Graduate School of Science, The University of Tokyo, 2-21-1 Osawa, Mitaka, Tokyo 181-0015, Japan}
\altaffiltext{19}{Faculty of Education, Saitama University, 255 Shimo-Okubo, Sakura, Saitama, 338-8570, Japan}




\begin{abstract}

  We report observations of the Type Iax supernova (SN Iax) 2012Z 
 at optical and near-infrared wavelengths from immediately after the 
 explosion until $\sim$ $260$ days after the maximum luminosity 
 using the Optical and Infrared Synergetic Telescopes for Education and 
 Research (OISTER) Target-of-Opportunity (ToO) program and the Subaru 
 telescope. 
 We found that the near-infrared (NIR) light curve evolutions 
 and color evolutions are similar to those of SNe Iax 2005hk and 2008ha. 
 The NIR absolute magnitudes ($M_{J}\sim-18.1$ mag and $M_{H}\sim-18.3$ mag) 
 and the rate of decline of the light curve ($\Delta$ $m_{15}$($B$)$=1.6 \pm 0.1$ mag) 
 are very similar 
 to those of SN 2005hk ($M_{J}\sim-17.7$ mag, $M_{H}\sim$$-18.0$ mag, and 
 $\Delta$ $m_{15}$($B$)$\sim1.6$ mag),
 yet differ significantly from SNe 2008ha and 2010ae ($M_{J}\sim-14 - -15$ mag and 
 $\Delta$ $m_{15}$($B$)$\sim2.4-2.7$ mag).
 The estimated rise time is $12.0 \pm 3.0$ days, which is significantly shorter 
 than that of SN 2005hk or any other Ia SNe.
 The rapid rise indicates that the $^{56}$Ni distribution may extend 
 into the outer layer  or that the effective 
 opacity may be lower than that in normal SNe Ia. 
 The late-phase spectrum exhibits broader 
 emission lines than those of SN 2005hk by a factor of 6--8. 
 Such high velocities of the emission lines indicate that the 
 density profile of the inner ejecta extends more 
 than that of SN 2005hk. 
  We argue that the most favored explosion scenario is
 a `failed deflagration' model, 
 although the pulsational delayed detonations is not excluded.

\end{abstract}

\keywords{supernovae: general --- supernovae: individual (SN~2012Z) 
--- supernovae: individual (SNe~2005hk)}

\section{Introduction}
  Type Ia supernovae (SNe Ia) provide a strong 
 constraint on the cosmological parameters \citep{Riess1998,Perlmutter1999}
 because their peak luminosities are well correlated with the widths of their 
 light curves \citep{Pskovskii1984,Phillips1993,Phillips1999,
 Altavilla2004,XWang2005,XWang2006,Prieto2006,Jha2007,Conley2008,Folatelli2010}.
 More luminous objects typically have slower rates of decline of their light curves.
 However, although normal SNe Ia exhibit such homogeneous properties,
 various diverse properties have recently been discovered, e.g., SNe Iax 
 \citep{Li2003,Phillips2007,Valenti2009}, super-Chandrasekhar SNe
 \citep{Howell2006,Hicken2007,Yamanaka2009a},
 and SNe Ia/IIn \citep{Hamuy2003,Aldering2006}. 

 SNe Iax are one of the most mysterious 
 SNe \citep{Li2003,Phillips2007,Sahu2008,Valenti2009,Foley2010a,Foley2010b,
 Narayan2011,Stritzinger2014a,White2015}. 
 The peak luminosities are significantly fainter than those expected 
 from the rates of decline of their light curves. 
 Their light curves do not exhibit a secondary
 maximum in the I band 20--30 days after the B-band maximum,
 and their spectra exhibit a blue continuum with absorption lines for 
 ions in relatively
 high ionization states during the early phases, e.g., 
 Fe~{\sc iii} rather than Si~{\sc ii} \citep{Branch2004}. The line velocity exhibits
 a diversity from 2000 to 8000 km~s$^{-1}$ among SNe Iax. These 
 velocities are significantly lower than those of normal SNe Ia \citep{Foley2013}.

  There is no comprehensive explosion model to
 describe the observed properties of SNe Iax.
 The small line velocities suggest that the explosion energy 
 could be much smaller than for normal SNe Ia.
 \cite{Phillips2007} proposed a pure deflagration model 
 \citep{Nomoto1984}
 to describe these observed properties. 
  With a pure deflagration model, strong
 mixing is expected to occur due to hydrodynamic instabilities
 \citep{Gamezo2003}. Such mixing could result in a singly peaked
 light curve in the near-infrared (NIR) \citep{Blinnikov2006}. 
 Alternatively, a core collapse scenario has also been proposed 
 to describe extremely faint SNe Iax \citep{Moriya2010}.

  A progenitor candidate of SNe Iax has been searched for in
 pre-explosion images \citep{Foley2010b,McCully2014}. 
 For SN 2012Z, a blue source was detected in the pre-explosion 
 images taken using the {\it Hubble Space Telescope} ({\it HST}) 
 \citep{McCully2014}. 
 The luminosity ($M_{V}\sim-5.3$ mag) and color were similar to those of an He star or 
 a quiescent phase of the helium nova V445 Pup \citep{Kamath2002,Iijima2008}.
 Although He~{\sc i} emission lines have been reported for a few SNe Iax, 
 spectra of SNe Iax do not necessarily exhibit He features 
 \citep{Jha2007,Foley2013}. 

  SN 2012Z was discovered at an unfiltered magnitude of 17.6 mag on Jan 
 29.15 UT ($t=-10.65$ d \footnote{We adopt 
  MJD $55965.8 \pm 3.0$ as $t=0$
 throughout this paper; see \S 3.1.}) by Lick Observatory Supernova Search 
 \citep[LOSS;][]{Filippenko2001} in the nearby galaxy NGC 1309 
 \citep{Cenko2012a}.
 \citet{Cenko2012b} reported a follow-up photometric 
 observation three minutes after discovery. 
 They found that the magnitude of SN 2012Z was
 $V=18.00\pm0.16$ mag.
 The spectrum was very similar to that of the luminous-class 
 SN Ia 1991T or the well-studied SN Iax 2005hk \citep{Cenko2012c,Meyer2012}. 
 Table 1 lists a summary of the parameters of SN 2012Z.

 In this paper, we describe continuous photometric observations in NIR 
 bands as well as optical ones. 
 Such wide-wavelength observations of SNe Iax 
 are rare \citep{Phillips2007,Stritzinger2014a}.
 We focus on the NIR properties and the rise time; 
 in particular, we show a detailed comparison with the bright SN 
 Iax 2005hk 
 \citep[$M_{V}\sim-18.1$ mag; ][]{Phillips2007,Sahu2008}, as well as 
 the faint SNe Iax 2008ha \citep[$M_{V}\sim-14.2$ mag; ][]{Valenti2009,Foley2010a} 
 and 2010ae \citep[$M_{V}\sim-14.5$ mag; ][]{Stritzinger2014a}.
 Explosion models are also discussed by comparing the expected outcomes
 with our observations.

\section{Observations and data reduction}

  Observations of SN 2012Z were performed in the framework of
 a Target-of-Opportunity program of the Optical and Infrared
 Synergetic Telescopes for Education and Research (OISTER),
 which utilizes small ground-based telescopes in Japan,
 South Africa, and Chile, as part of an inter-university collaboration.
 The aim of OISTER is to investigate various transient sources and 
 variable stars, including $\gamma$-ray bursts, active galactic nuclei, supernovae, 
 and cataclysmic variables. 
 Table 2 summarizes the instruments, telescopes, and
 observatories used for the observations of SN 2012Z.


 \subsection{NIR photometric observations}

  We performed $J$, $H$, and $K_{s}$-band photometric observations
 using the Wide-Field Camera \citep[WFC; ][]{Yanagisawa2014} installed on the 0.91-m
 telescope at the Okayama Astrophysical Observatory (OAO) on five nights;
 the Simultaneous three-color InfraRed Imager for Unbiased
 Survey \citep[SIRIUS; ][]{Nagayama2003} installed at the
 1.4-m InfraRed Survey Facility (IRSF) telescopes at the South
 Africa Astronomical Observatory (SAAO) on 20 nights;
 the Nishi-harima Infrared Camera (NIC) installed at the 
 Cassegrain focus of the 2.0-m Nayuta telescope at the 
 Nishi-Harima Astronomical Observatory on eight nights;
 the Infrared Imager and Spectrograph \citep[ISLE; ][]{Yanagisawa2006} 
 installed at the Cassegrain focus of the 1.88-m telescope
 on one night; and the InfraRed Camera (IRC) installed at 
 the 1.0-m telescope at the Iriki Observatory on 12 nights.

  The obtained data were reduced as follows according to a standard procedure 
 for ground-based NIR photometry. The sky background was subtracted 
 using the template sky image derived in each dithering observation set.
 We performed point-spread function (PSF)
 fitting photometry using the $DAOPHOT$ package 
 of $IRAF$. 

  \subsection{Optical photometric observations}

  In addition to the NIR photometric observations, we performed
 $B$, $V$, $R$, and $I$-band photometric observations using 
 the Hiroshima One-shot Wide-field Polarimeter 
 \citep[HOWPol; ][]{Kawabata2008} 
 installed at the 1.5-m Kanata telescope on 26 nights, and
 $U$, $B$, $V$, $R$, and $I$-band observations with 
 the Multi-Spectral Imager \citep[MSI; ][]{Watanabe2012} 
 installed at the 1.6-m Pirka telescope
 on four nights.
 We also performed $g'$, $R$, and $I$-band observations using 
 three robotic
 observations systems: the Multicolor Imaging Telescopes for
 Survey and Monstrous Explosions \citep[MITSuME; ][]{Kotani2005}; 
 with MITSuME at OAO \citep{Yanagisawa2010} on 14 nights; at the Akeno 
 Observatory on 10 nights; and at the 
 Ishigaki-jima Astronomical Observatory on seven nights. 
 Adding to the OISTER collaboration, we performed
 $B$, $V$, $R$, and $I$-band photometric observations using 
 the 1.88-m telescope at the Kottamia Observatory on two nights.
 These data were reduced according to a standard procedure
 for charge-coupled device (CCD) photometry. We used PSF photometry, as with 
 the NIR data.
 
\begin{figure*}
  \begin{center}
    \begin{tabular}{c}
\resizebox{100mm}{!}{\includegraphics{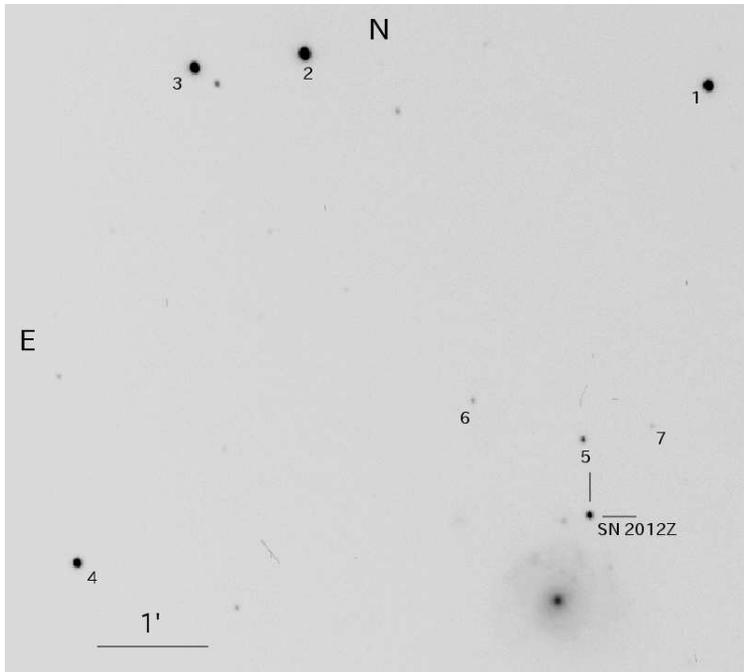}} 
    \end{tabular}
    \caption{The $I$-band image with SN 2012Z obtained using HOWPol attached to the
    Kanata telescope on 2012 Feb 12 ($t=+4.1$ d). }
    \label{fov}
  \end{center}
\end{figure*}

 
\subsection{Optical spectroscopy}

  We performed optical spectroscopic monitoring
 using HOWPol on 12 nights, and with the low-resolution spectrograph
 LOSA/F2 \citep{Shinnaka2013} mounted on the Nasmyth stage
 of the 1.3-m Araki telescope at the Koyama Astronomical Observatory
 on four nights (Table 3).
 The wavelength coverage of HOWPol was 
 4500--9000 \AA\ and the spectral resolution was $R =$ $\lambda$/$\Delta$$\lambda$$\simeq$ 
 400 at 6000 \AA; the wavelength coverage of LOSA/F2 was 4200--8000 \AA\, with a spectral resolution of $R$ $\simeq$ 600 at 6000 \AA.
 During the data reduction procedure, the wavelengths were calibrated
 using telluric emission lines acquired in the object frames for 
 the HOWPol data, and using the FeArNe arc lamp frames for 
 the LOSA/F2 data. The fluxes were calibrated using the frames of 
 spectrophotometric standard stars taken on the same night as the 
 object data.
 We removed the strong telluric absorption features from the object
 spectra using the spectra of high-temperature standard stars.

 \subsection{Late phase observations}

  We performed spectroscopic observations of SN 2012Z using
 the Faint Object Camera and Spectrograph \citep[FOCAS; ][]{Kashikawa2002} 
 attached to the Cassegrain focus of the 8.2-m
 Subaru telescope on Oct 23 UT ($t=+261.3$ 
 d). For the shorter wavelength observations, we used a B300 grism. 
 For the longer wavelengths, we used an R300 grism and an O58 
 order-cut filter. The composite wavelength coverage was 4600--9600 \AA.
 We used the 0$''$.8-width slit; the 
 resulting wavelength resolution was $R\simeq660$ at 6000 \AA. 
 During the data reduction procedure, the wavelength was calibrated
 using the frames of the ThAr arc lamp. We also obtained
 $B$, $V$, $R$, and $I$-band imaging observations for the SN and 
 photometric standard stars.

 \subsection{Photometric calibration}
 
  We performed relative photometry using the local reference stars 
 in the field of SN 2012Z (see Figure \ref{fov}).
 For the NIR data, we used the magnitudes from the 2MASS catalog 
 for the references stars \citep{Persson1998}. 
  We adopted a square average of the standard deviation of
 the SN and the systematic error of the reference star
 magnitudes for the observational errors. A summary of the results of
 the NIR photometry is listed in Table 4.
 
  To calibrate the $B$, $V$, $R$, and $I$-band magnitudes of 
 the reference stars, we used the photometric data of standard stars 
 in the SA95 region \citep{Landolt1992}, which were obtained using 
 HOWPol on a photometric night.
 We performed relative photometry for the SN commonly using these
 magnitudes, except for the data of MITSuME at OAO.
 For the data of MITSuME at OAO, the $g'$, $R$, and $I$-band data 
 were calibrated using the SA96 \citep{Landolt1992} region obtained 
 using the instrument itself. For the $U$-band data
 at Pirka, we obtained the magnitude of the references stars
 using extrapolation with the $U-B$ versus $B-V$ relation of
 the Landolt standard sequence \citep{Landolt1992}; therefore,
 the observational error was relatively large. Table 8 lists a summary of the optical photometry data. 

 Table 5 lists a summary of the obtained magnitudes of
 the references stars. Note that we reduced the
 systematic error among the various instruments 
 using a color-term correction in a consistent manner, and included
 newly performed observations of the standard stars in the M67
 \citep{Stetson1987} and SA98 \citep{Landolt1992} regions.

\section{Results}
\subsection{Light curves}
   
\begin{figure*}
  \begin{center}
    \begin{tabular}{c}
      \resizebox{160mm}{!}{\includegraphics{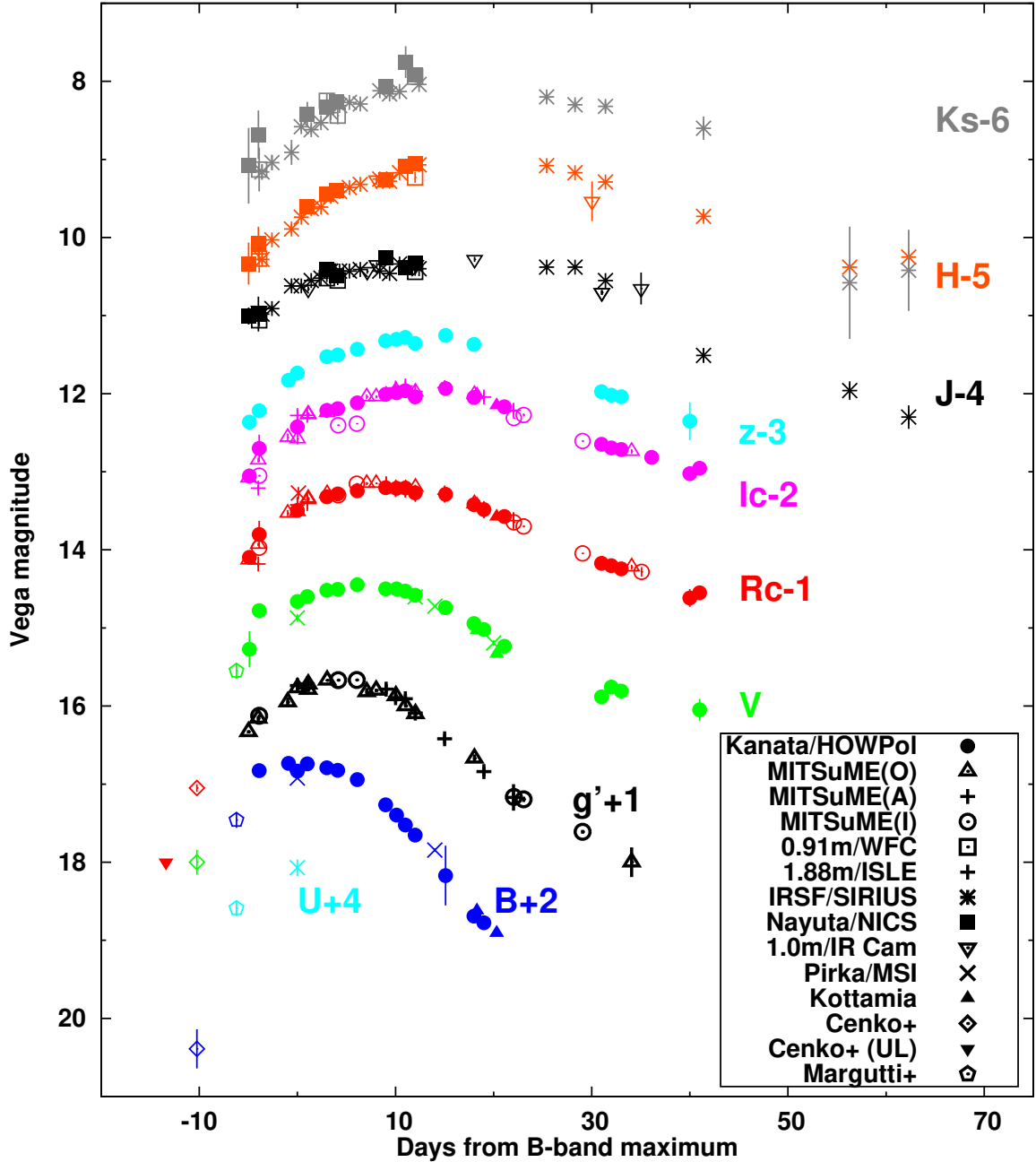}} \\
    \end{tabular}
    \caption{$U$, $B$, $g'$, $V$, $R$, $I$, $z'+Y$, $J$, $H$, and $K_{s}$-band light curves
 of SN 2012Z. Galactic extinctions were corrected for. The host galactic extinction was 
 negligible. The different colors denote different-band filters, and the shape of the symbols corresponds to the instrument used (see the figure legend). The open-diamond symbols denote the magnitudes
 reported by \citet{Cenko2012a}. The open-pentagon symbols denote the 
 $U$, $B$, and $V$-band magnitudes obtained by 
 $Swfit$/UVOT \citep{Cenko2012b, Margutti2012}.} 
    \label{epl}
  \end{center}
\end{figure*}

 Figure \ref{epl} shows the $U$, $B$, $g'$, $V$, $R$, $I$, $z'+Y$, $J$, $H$, and $K_{s}$-band
 light curves. Galactic extinctions were
 corrected for using the color excesses $E$($B-V$)$=0.036$ mag 
 \citep{Schlafly2011}. We assumed negligible host 
 galactic extinction by comparing the color evolutions
 with those of the same subclass, including the near-infrared photometry.
 
  We found that the $B$-band maximum magnitude 
 was $14.74 \pm 0.01$ mag and the date was 
  MJD $55965.8 \pm 3.0$. The peak date was 1.3 days earlier 
 than that
 reported  by \citet{Stritzinger2015}.  
 This date was assumed as $t=0$ throughout	
 this paper. We also evaluated the maximum magnitudes and dates in all other 
 bands (see Table 6). 
 As with other SNe Iax, the NIR light curves for SN 2012Z 
 exhibited a single maximum. The maxima for redder 
 bands appeared later.
  For example, the $K_{s}$-band maximum occurred 
  at $t=+16$ d, which was significantly later than 
  that of a typical SN Ia \citep[$t=-$3 d][]{Krisciunas2003}.
 We calculated the rate of decline of the $B$-band light curve as 
 $\Delta$ $m_{15}$($B$)$=1.57 \pm 0.07$ mag, which is slightly
 larger than that reported by \citet{Stritzinger2015}. 
 We also calculated $\Delta m_{15}$($\lambda$) of the
 $V$, $R$, $I$, $J$, $H$, and $K_{s}$-band light curves
 (see Table 6). We found that the rates of decline 
 were smaller in the redder bands, which corresponds to the wider shapes,
 except for the $J$-band.

\begin{figure*}
  \begin{center}
    \begin{tabular}{c}
      \resizebox{90mm}{!}{\includegraphics{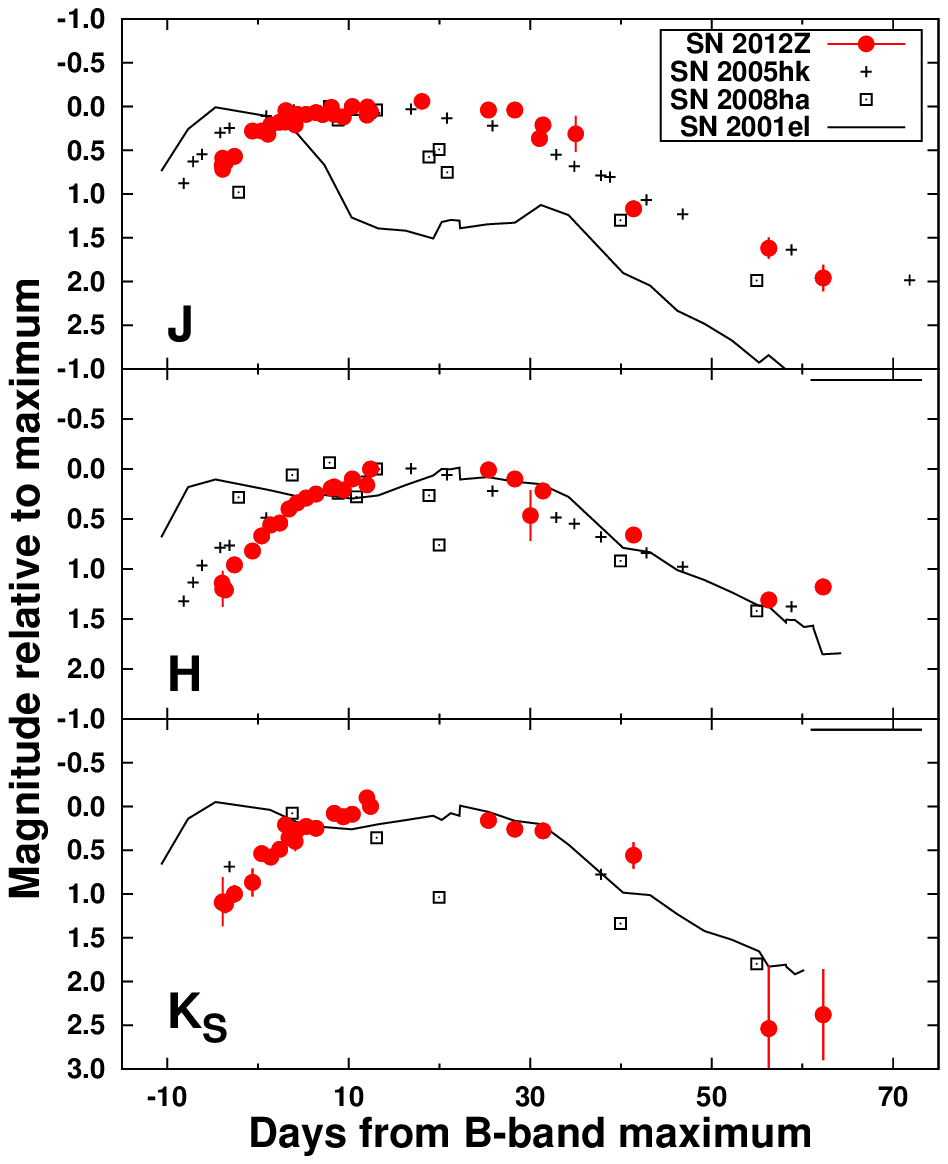}} 
    \end{tabular}
    \caption{$J$, $H$, and $K_{s}$-band light curves compared with those of 
    SNe 2005hk ($\Delta m_{15}$($B$)$=1.7\pm 0.1$ mag; \cite{Phillips2007}), 
   2008ha ($\Delta m_{15}$($B$)$=2.17 \pm 0.02$ mag; \cite{Foley2010a}) and 2001el 
    ($\Delta m_{15}$($B$)$=1.13 \pm 0.04$ mag; \cite{Krisciunas2003})
    in the range of $t=-15$ to $+75$ d. The maximum magnitudes were 
    shifted to zero magnitudes for the sake of comparison.}
    \label{nirlcc}
  \end{center}
 \end{figure*}

 Figure \ref{nirlcc} shows the $J$, $H$, and $K_{s}$-band light curves as well as 
 those of SNe Iax 2005hk \citep[$\Delta m_{15}$($B$)$=1.7\pm$0.1 mag;][]{Phillips2007}, 
 2008ha  \citep[$\Delta m_{15}$($B$)$=2.17 \pm 0.02$ mag;][]{Foley2010a} 
 and a normal SN 2001el 
 \citep[$\Delta m_{15}$($B$)$=1.13 \pm 0.04$ mag; ][]{Krisciunas2003}.
 Our near-infrared light curves exhibited the best time coverage
 among SNe Iax.

  NIR light curves of SN 2012Z and other SNe Iax commonly
 show a single peak, in contrast to the double peaks observed with normal Type 
 Ia SNe. For the $J$-band light curves, the overall shape was 
 similar to that of SN 2005hk (at $t=-5 - +60$ d).
 A small difference was observed in the rise time; i.e.,
 SN 2012Z exhibited a faster rise than did SN 2005hk.
 SN 2008ha also exhibited a faster decline than SN 2012Z.
 The magnitude (relative to the maximum) of SN 2008ha
 was fainter than that of SN 2012Z by $0.5-1.0$ mag at $t=+20$ d.
 Similar trends were also found in the $H$-band light curves; i.e.,
 SN 2012Z exhibited a faster rise, and SN 2008ha a faster decline.
 For the $K_{s}$ band, dense data are not available for other 
 SNe Iax. Two photometric points of SN 2005hk are consistent 
 with those of SN 2012Z.


  In summary, the near-infrared light curves were
  highly similar to those of SN 2005hk among $t=-5-+60$ d,
  except during the rise,
  as determined from the optical bands \citep{Stritzinger2015}. 

\begin{figure*}
  \begin{center}
    \begin{tabular}{c}
      \resizebox{90mm}{!}{\includegraphics{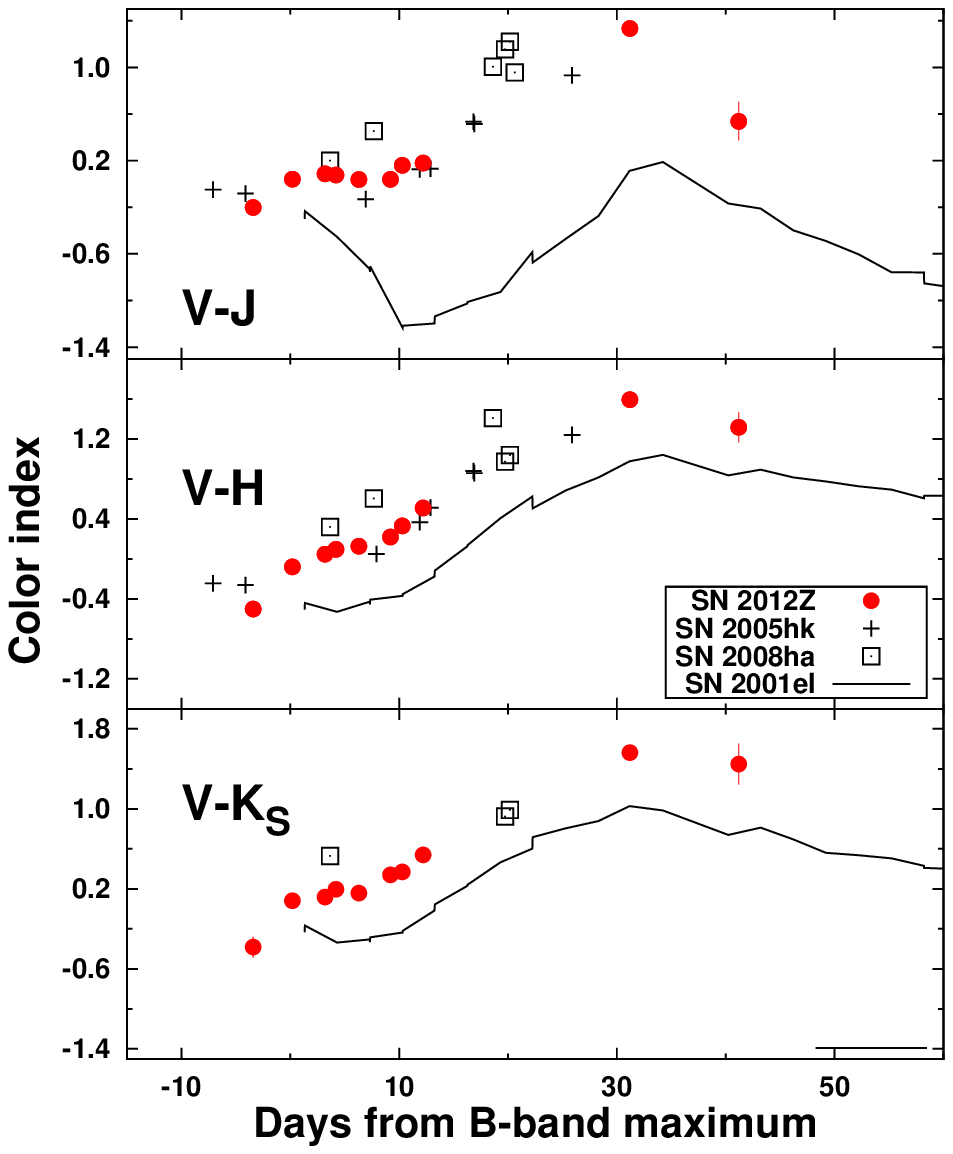}} 
    \end{tabular}
    \caption{$V-J$, $V-H$, and $V-K_{s}$
    color evolutions plotted together with those of SNe 2005hk, 2008ha, and 2001el.
    Galactic and host galactic extinctions are corrected for.}
    \label{nircl}
  \end{center}
\end{figure*}

 Figure \ref{nircl} shows the color indices $V-J$, $V-H$, and $V-K_{s}$ 
 of SN 2012Z. 
 These data were corrected for Galactic extinctions only.
 The colors continuously evolved toward the red during our observations. 
 The $V-J$ and $V-H$ evolutions of SN 2012Z
 were similar to those of SN Iax 2005hk.
 On the other hand, the color evolutions differed 
 significantly from those of typical SNe Ia.
 In particular, the $V-J$ color of SNe Iax exhibited monotonic behavior,
 evolving toward redder colors with time, whereas a typical SN 2001el
 exhibits non-monotonic behavior, which results from the
 double-peak light curve in the $J$ band.

  Because the near-infrared flux is not attenuated significantly due to dust, 
 the optical to near-infrared wavelengths enable us to estimate
 the extinction accurately if the intrinsic color is known \citep{Krisciunas2004}.
 Because the absolute color was 
 consistent with that of the extinction-corrected SN 2005hk 
 \citep[$E$($B-V$)$=0.09$ mag; ][]{Phillips2007},
 the extinction of SN 2012Z may be considered negligible.
 The $V-J$ and $V-H$ colors of SN 2012Z were bluer than those of SN 2008ha,
 which may result from uncertainties in the color 
 excess \citep{Foley2013}.

 The color of the NIR bands was similar to those of SNe Iax, 
 as shown in Figure 5. The $J-H$ and $J-K_{s}$ color evolutions exhibited
 a monotonous increase.
 The $J-K_{s}$ color 
 evolution was similar to those of SNe 2005hk and 2008ha, 
 despite the sparse data. 
 The $H-K_{s}$ color was similar to those of SNe 2005hk 
 and 2008ha, although these were sparse.
 Only the $H-K_{s}$ color exhibited a similarity between SN 2012Z 
 and normal SNe Ia (see the bottom panel of Figure 5).
 Such homogeneity in the $H-K_{s}$ is interesting,
 whereas all the other colors ($V-$NIR colors, $J-H$ 
 and $J-K_{s}$ colors) exhibited a large difference between 
 other SNe Iax and normal SNe Ia.


\begin{figure*}
  \begin{center}
    \begin{tabular}{c}
      \resizebox{90mm}{!}{\includegraphics{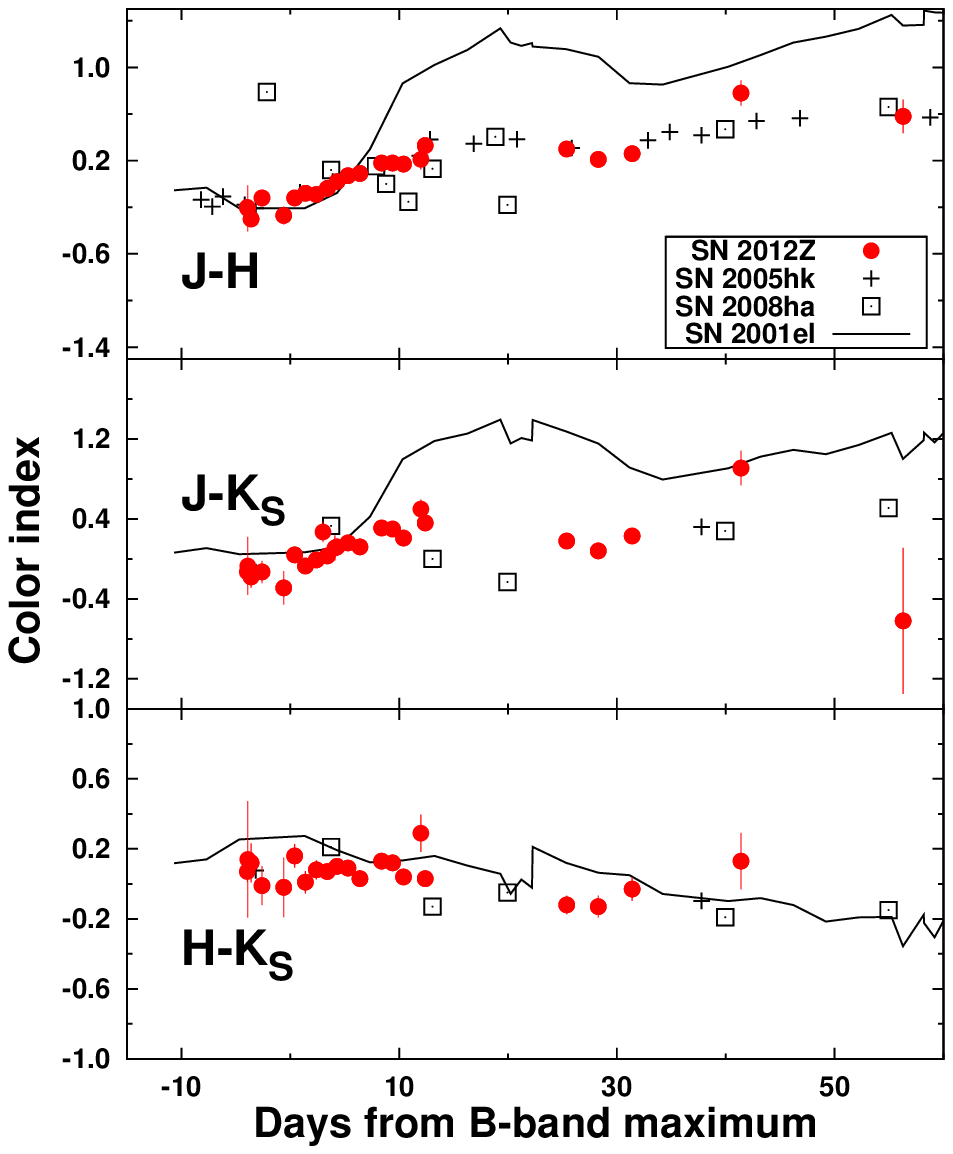}} 
    \end{tabular}
    \caption{Same as Figure 4, but for NIR-NIR color evolutions.} 
    \label{nircl2}
  \end{center}
\end{figure*}

 \subsection{Rise time}
 

  The rise times 
 of SNe Ia have been studied statistically
 \citep{Hayden2010,Ganeshalingam2011,Gonzalez2012} and individually
 \citep{Foley2012,Silverman2012a,Nugent2011,Bloom2012,Yamanaka2014,
 Zheng2013,Zheng2014,Goobar2014}.

\begin{figure*}
  \begin{center}
    \begin{tabular}{c}
      \resizebox{50mm}{!}{\includegraphics{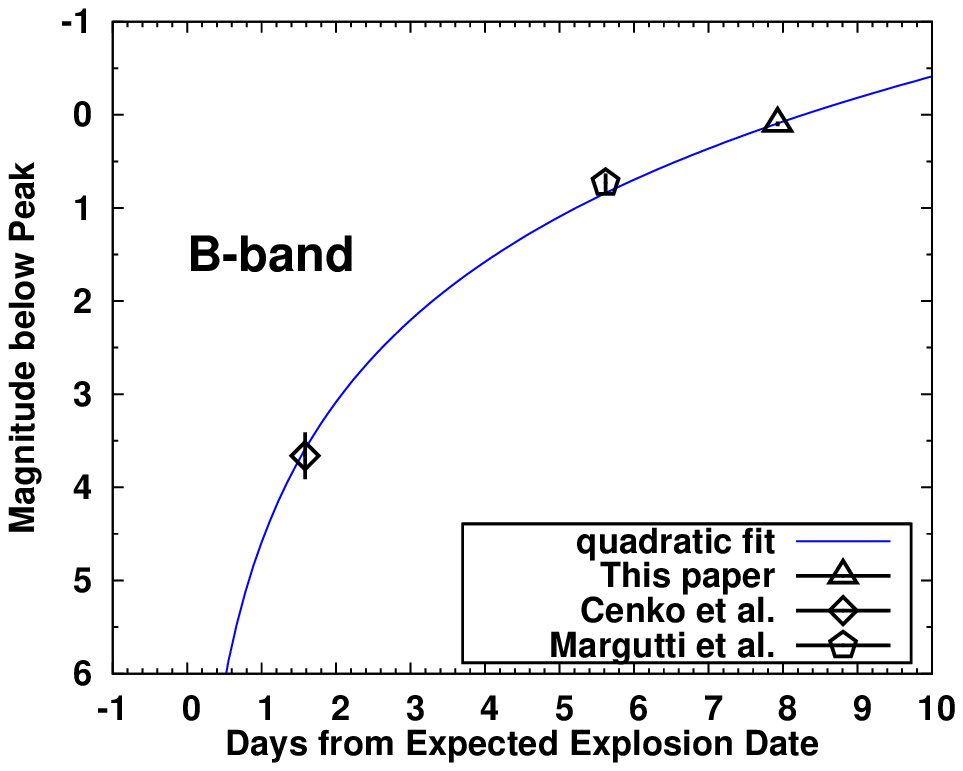}} 
      \resizebox{50mm}{!}{\includegraphics{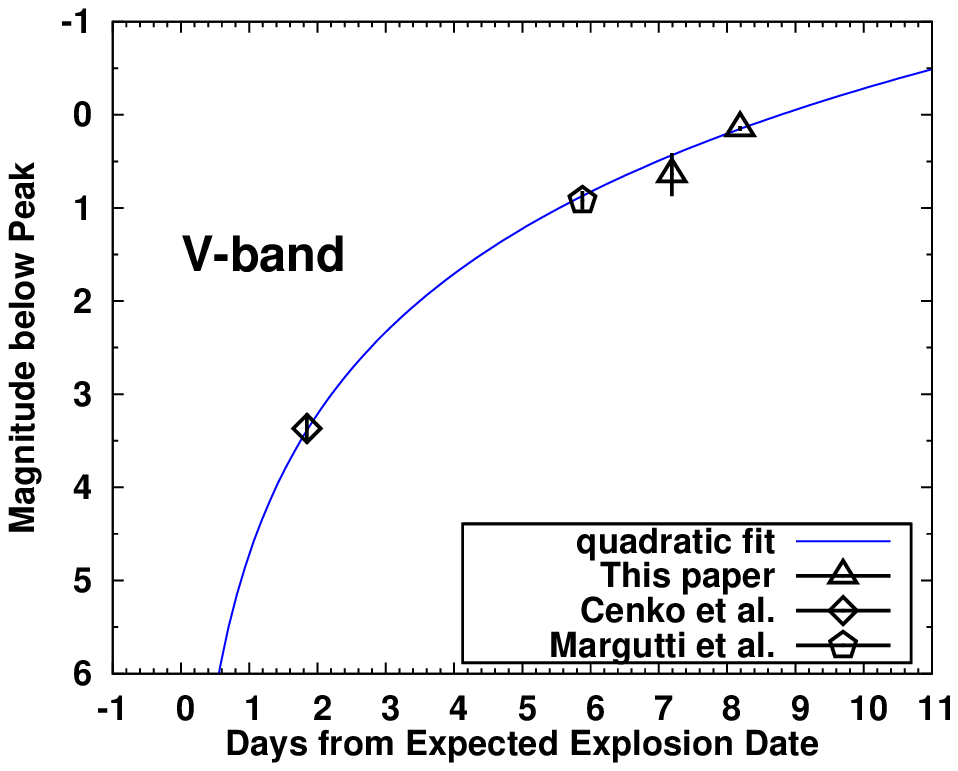}} 
        \resizebox{50mm}{!}{\includegraphics{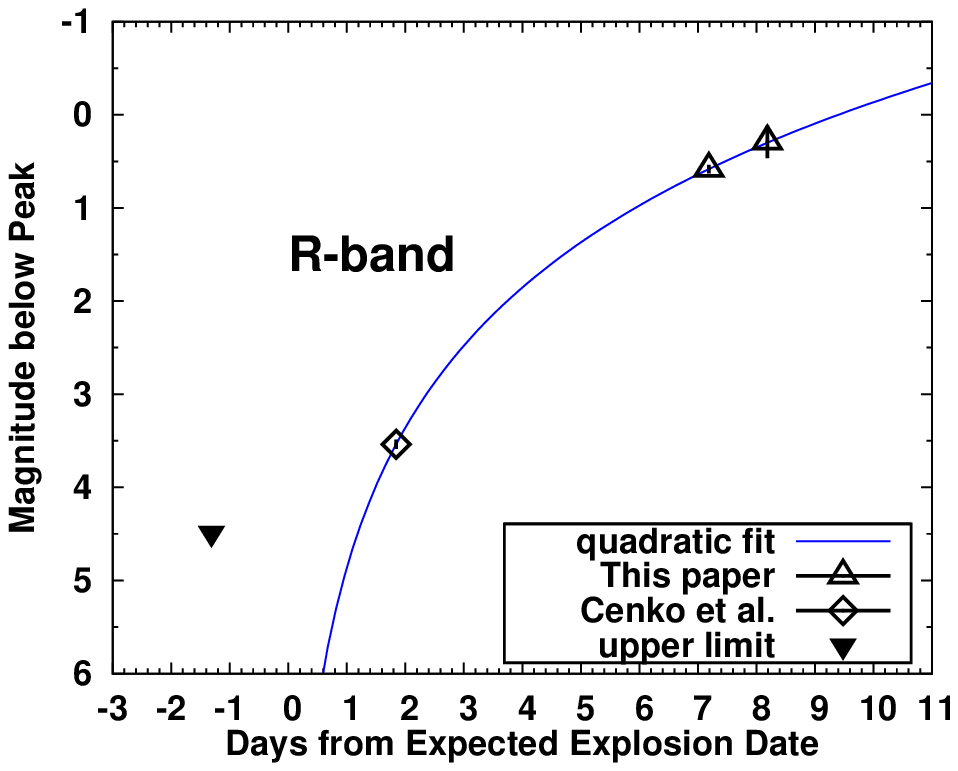}} 
    \end{tabular}
    \caption{$B$, $V$, and $R$-band rise curves of SN 2012Z with 
    quadratic fits (shown by blue doted curves). 
    These data include the $b$ and $v$ magnitudes obtained using the 
    $Swift/UVOT$ telescope \citep{Margutti2012}, and the immediately 
    observed $B$, $V$, and $R$-band data obtained using 
    LOSS \citep{Cenko2012a} and HOWPol. 
    The  filled triangle symbols in the right panel denote the upper-limit 
    magnitudes reported by \citet{Cenko2012b}. The maximum magnitudes were
    shifted to zero.}
    \label{rt}
  \end{center}
\end{figure*}

  SN 2012Z exhibited rapid rises in the $B$, $V$, and $R$ bands
 following its discovery \citep{Cenko2012b},
 but was not detected at 3.2 days
 before discovery \citep{Cenko2012b}.
An upper limit of the 
 magnitude was given as $R>19.0$ mag on Jan 26 ($t\sim-13$ d).  
 We fitted the published data and 
 our data in the $B$, $V$, and $R$-band using a quadratic function, 
 as shown in Figure \ref{rt}.


  We calculated the rise times of SN 2012Z to be $11.8$, $12.1$, and $12.1$ days 
 in the $B$, $V$, and $R$ bands, respectively.
 The average rise time was  $t_{rise}=12.0 \pm 3.0$
 days, and 
 the explosion date was MJD 55953.4. 
 The rises of the light curves were well 
 described using single quadratic functions. 
 The smoothly rising curves indicate that there were no 
 strong interactions between the ejecta and companion stars or
 dense circumstellar materials (CSMs). We discuss possible progenitor 
 systems of SN 2012Z in \S 4.2.

 \subsection{Spectra}

\begin{figure*}
  \begin{center}
    \begin{tabular}{c}
      \resizebox{90mm}{!}{\includegraphics{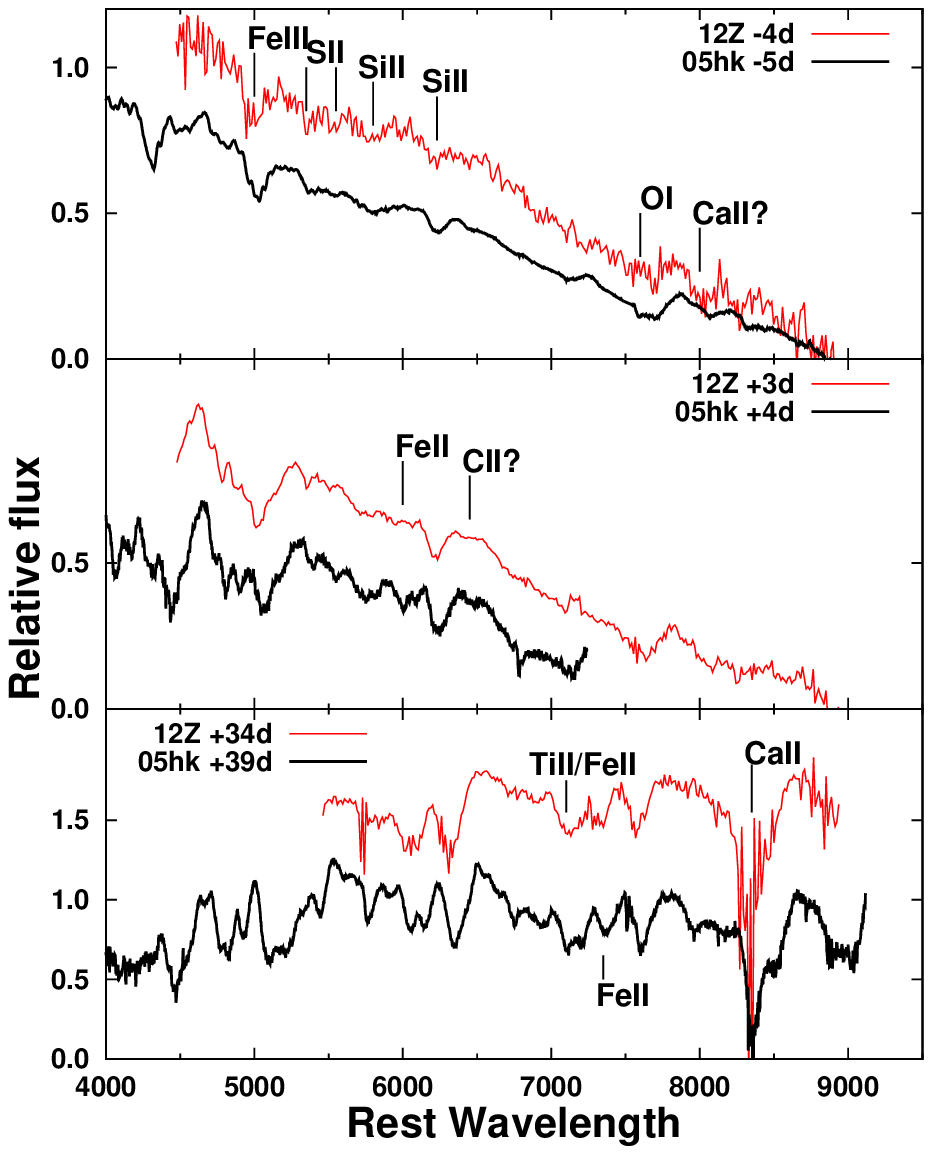}} 
    \end{tabular}
    \caption{(Top panel) Spectrum of SN 2012Z at $t=-4$ d compared with that of
    Type 
    Iax SN 2005hk \citep{Phillips2007} at $t=-5$ d. The wavelength was calibrated to the rest frame of
    each host galaxy. The spectrum exhibited Si~{\sc ii}$\lambda$6355, 
    Si~{\sc ii}$\lambda$5972 and S~{\sc ii} W-shape features as well as 
    Fe~{\sc iii} $\lambda$ 5267 absorption lines. The Ca~{\sc ii}~IR triplet and 
    O~{\sc i}~$\lambda$7774 were also detected. (Middle panel) Same as
    the top panel, but at $t=+3$ d. Possible C~{\sc ii}~$\lambda$6580 absorption 
    lines were observed. (Bottom panel) Same as the top panel, but at $t=34$d.
    Many Fe~{\sc ii} absorption lines were clearly visible. }
    \label{spce}
  \end{center}
\end{figure*}

  The spectra were compared with those of SN Iax 2005hk
 \citep{Phillips2007}, as shown in Figure \ref{spce}. The samples of
 SN 2005hk were collected from the SUSPECT \footnote{http://www.nhn.ou.edu/~suspect/} 
 and WISEREP \footnote{http://wiserep.weizmann.ac.il/} databases.

  At $t=-5$ d, the spectral
 features were similar to those of SN 2005hk, except for the
 slightly larger blueshift of all features and the shallower depth.
 The spectral features at $t=+4$ d also exhibit a slightly larger blueshift
 than those of SN 2005hk. 
 The C~{\sc ii} absorption line was barely observable between $t=+1$ and
 $t=+12$ d, which may also be a common property with SN 2005hk. 
  The line velocity was $\sim$7000 km~s$^{-1}$ between 
 $t=+1$ and $+12$ d. 
 Also note that C~{\sc ii} was observed only
 during the post-maximum phase.

\begin{figure*}
  \begin{center}
    \begin{tabular}{c}
      \resizebox{90mm}{!}{\includegraphics{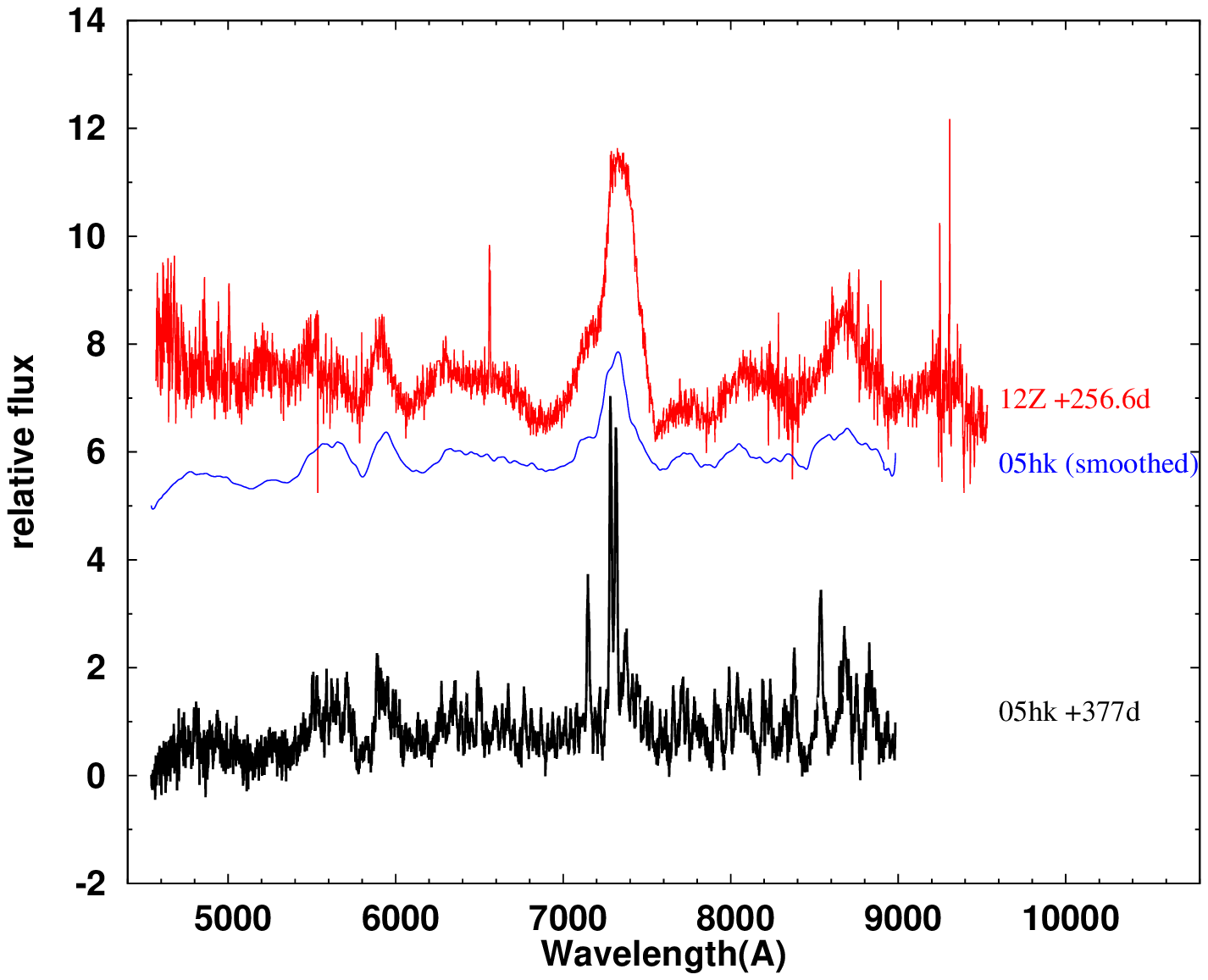}} 
    \end{tabular}
    \caption{Nebular spectrum of SN 2012Z at $t=256.6$ d (red curve) and the 
spectrum of SN 2005hk \citep{Sahu2008} at $t=377$ d (black curve). 
The blue curve denotes the artificially broadened
      spectrum of SN 2005hk. }
    \label{spcl}
  \end{center}
\end{figure*}

 The spectrum of SN 2012Z at $t=+261$ d is shown in Figure \ref{spcl}, 
 along with that of SN 2005hk at $t=+377$ d \citep{Sahu2008}.
 The spectral profiles differed from those of SN 2005hk.
 Strong emission features were observed around 7200 \AA. This 
 feature is blended with the [Ca~{\sc ii}] $\lambda$$\lambda$7291, 7323 
 and [Fe~{\sc ii}] $\lambda$7155 emission lines. 
 The emission from the permitted transitions 
 (the Na~{\sc i} D and Ca~{\sc ii} IR triplet) were also clearly visible.
 These features were observed in the spectrum of SN 2005hk;
 however, the linewidths were significantly broader than those of SN 2005hk.
 The full-widths at half-maximum (FWHM) values were estimated as 
 6300, 6200, 5600, and 6300 km~s$^{-1}$
 for the [Ca~{\sc ii}], [Fe~{\sc ii}], Na~{\sc i} D, and Ca~{\sc ii} IR
 triplet lines, respectively. The velocities of those 
 lines were approximately 800--1000km~s$^{-1}$ for SN 2005hk \citep{Sahu2008}. 
 Such broad emission lines have not been observed in other SNe Iax 
 \citep{Jha2007,Foley2013}.
 We discuss the inner structure of the ejecta in \S4.3.

\begin{figure*}
  \begin{center}
    \begin{tabular}{c}
      \resizebox{90mm}{!}{\includegraphics{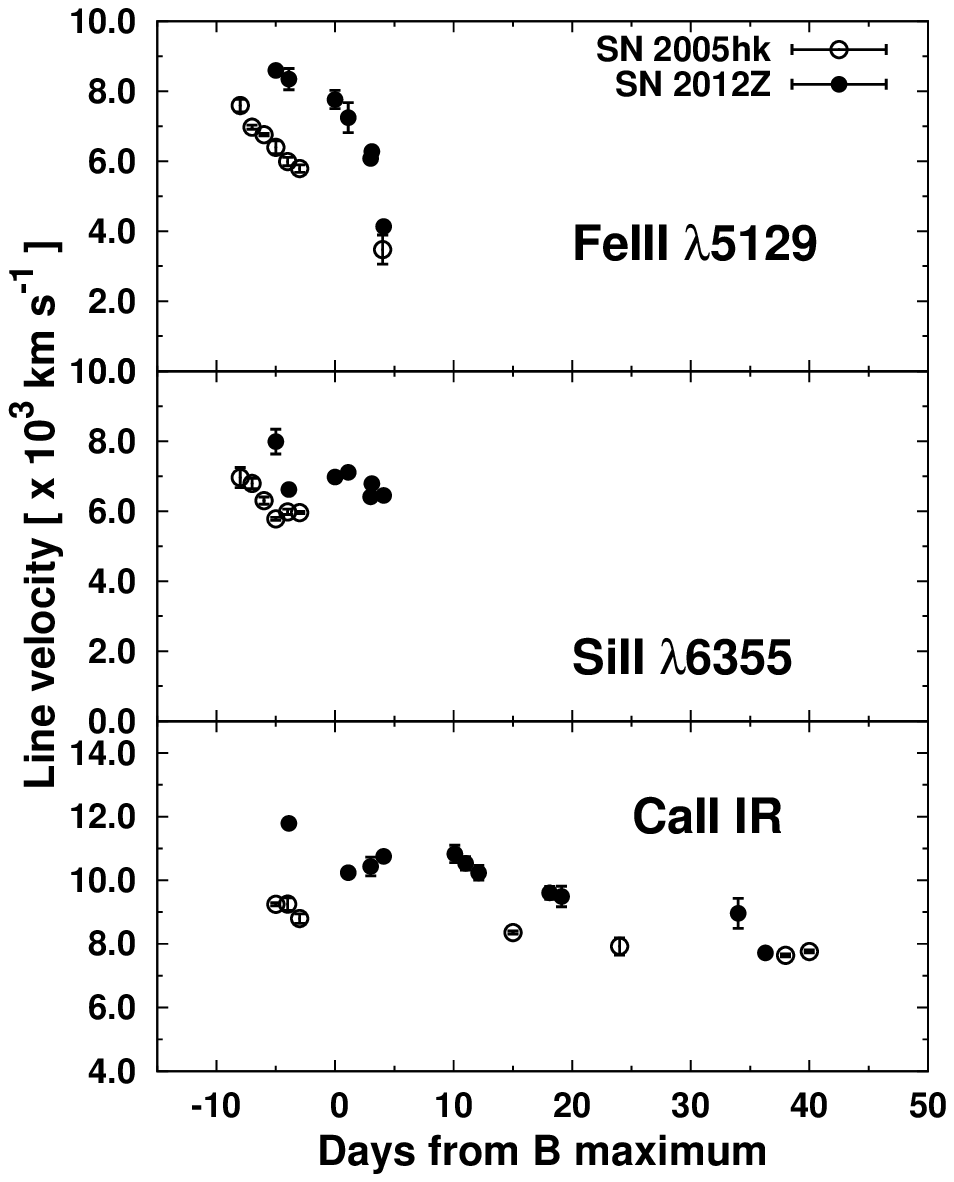}} 
    \end{tabular}
    \caption{(Top panel) Fe~{\sc iii}$\lambda$5129 line velocity 
    evolution of SN 2012Z (filled circles). This is compared 
    with that of SN 2005hk (open circles). 
    The horizontal axis is days from the $B$-band maximum. 
    (Middle panel) Same as the top panel, 
    but the Si~{\sc iii}$\lambda$6355 lines are shown. (Bottom panel)
    Same as the top panel, 
    but the Ca~{\sc ii}~IR~triplet lines are shown.}
    \label{lv}
  \end{center}
\end{figure*}

  We measured the line velocities of SNe 2012Z and 2005hk. 
 The velocities were estimated from the absorption minima.
 The recession velocities of the host galaxies
 were corrected for, according to the values from $NED$ 
 \footnote{NASA/IPAC Extragalactic Database; http://ned.ipac.caltech.edu/}. 
 The line velocities of Si~{\sc ii} $\lambda$6355, Fe~{\sc iii} $\lambda5129$,
 and Ca~{\sc ii} IR triplet of SN 2012Z
 are shown in Figure \ref{lv}, and are compared with those 
 of SN 2005hk in Figure 9.
 The rate of decline of the Si~{\sc ii} line velocity was
 $\sim$150 km~s$^{-1}$~d$^{-1}$ between $t=-5$ and $+10$ d.
 This is similar to that of SN 2005hk, and 
 as large as that of the high-velocity SNe Ia \citep{Benetti2005,Yamanaka2009b}.
 The velocity was 1500 km~s$^{-1}$ greater than that of SN 2005hk.  
 After $t=+12$ d, the Si~{\sc ii} $\lambda$6355 absorption was strongly
 blended by other lines, e.g., Fe~{\sc ii}
 \citep{Branch2004,Sahu2008}. The Fe~{\sc iii} line
 velocity was always 2000 km~s$^{-1}$ greater than that of SN 2005hk.
 We also show the line velocity of the
 Ca~{\sc ii} IR triplet in the bottom panel of Figure 9. The rate of decline 
 appeared to be flat compared 
 with those of Si~{\sc ii} and Fe~{\sc iii}. 
 A similar trend was also observed for SN 2005hk. 
 \citet{Stritzinger2015} found that
 the line velocities in Ca~{\sc ii} IR triplets and H\&K exhibit 
 flat evolutions.

\subsection{Environment}


 The environment of the host galaxies of SNe Iax has 
 been investigated based on morphological classifications
 \citep{Valenti2009,Foley2010b} and the strength of the 
 emissions from the H{\sc ii} regions \citep{Lyman2013}. 
 \citet{Lyman2013}
 reported that the distribution of H$\alpha$ intensities 
 of SN locations for SNe Iax is similar to that of core collapse 
 Type IIP SNe.
 The host galaxies
 of most SNe Iax are spiral galaxies, and include some star-forming 
 regions \citep[see also][]{Foley2010a, Valenti2009} (except for that
 of SN 2008ge \citep{Foley2010b}). The samples
 include SN 2012Z. 
 However, measurements of the metallicity of the explosion locations 
 are rare for SNe Iax.

 The FOCAS spectrum at $t=+261$ d exhibited weak emission 
 lines of H$\alpha$, [N~{\sc ii}]$\lambda$6583, H$\beta$, and 
 O~{\sc iii}$\lambda$5007. Using the O3N2 metallicity indicator 
 (i.e., the ratio of [N~{\sc ii}]/H$\alpha$ to 
 [O~{\sc iii}$\lambda$5007/H$\beta$) \citet{Pettini2004}),
 we estimate the metallicity as 12+log(O)$< 8.51 \pm 0.31$. 
 This is consistent with those of the previous two objects.
 These observations indicate that the environment of SN 2012Z might 
 be similar to those of SNe 2008ha and 2010ae 
 \citep{Foley2010a,Stritzinger2014a}.


 \subsection{Quasi-bolometric light curves}
  To determine the kinetic energy and ejecta mass of 
 SN 2012Z, we compare the quasi-bolometric light curve with
 an analytic model. First, we constructed quasi-bolometric
 light curves for SNe 2012Z and 2001el.
 The $B$, $V$, $R$, $I$, $J$, $H$, and $K_{s}$-band fluxes were 
 integrated using each filter passband function 
 \citep{Bessell1990,Bessell1998}. The ratio of 
 the fluxes of the NIR ($J$, $H$, and $K_{s}$ bands) to the 
 integrated fluxes ($B$, $V$, $R$, $I$, $J$, $H$, and $K_{s}$ bands) at 
 peak luminosity was estimated to be $\sim0.20$ for SN 2012Z, which is 
 larger than that of a normal SN Ia \citep[$\sim$0.1;][]{XWang2009}. 
 We assume that the $B$, $V$, $R$, $I$, $J$, $H$, and 
 $K_{s}$-band fluxes contribute $80\%$ of the total
 bolometric luminosity \citep{Stritzinger2006}. The rise time 
 was $t_{rise}=12.0$ days. 
 The $^{56}$Ni mass was estimated to be $\sim0.18M_{\odot}$, 
 which is considerably smaller than the average mass of SNe Ia 
 , although the uncertainty is relatively 
 large due to the uncertainty in the rise time.

  The ratio of the evolution of the NIR to the integrated flux 
 exhibited a monotonic increase from $+8$ to $+25$ days since the 
 explosion date. This differs significantly in comparison 
 with typical SNe Ia \citep{XWang2009}, but 
 is consistent with the 
 $V - NIR$ color evolutions (see \S 3.1). 

  The quasi-bolometric light curve was compared with an
 analytical model derived using \citet{Maeda2003} 
 \begin{equation}
 L(t)= M_{\rm 56Ni}((S_{\rm 56Ni}+S_{\rm 56Co})(1-e^{-\tau})+S_{\rm 56Co}{\cdot}f_{p}), 
 \end{equation}
 where $M_{\rm 56Ni}$ is the mass of the ejected $^{56}$Ni 
 and $f_{p}$ is the positron fraction.
 The energy release rates ($S_{\rm 56Ni}$ and $S_{\rm 56Co}$) 
 are expressed as 
 \begin{equation}
 S_{\rm 56Ni}=(3.90\times10^{10})e^{-t/t_{\rm 56Ni}} 
 {\rm erg~s}^{-1}~{\rm g}^{-1}
  \end{equation}
and
 \begin{equation}
 S_{\rm 56Co}=(7.01\times10^{9})(e^{-t/t_{\rm 56Co}} - e^{-t/t_{\rm 56Ni}})~~{\rm erg~s}^{-1}~{\rm g}^{-1}.
 \end{equation} 
 The decay timescales are $t_{\rm 56Ni}=8.8$ and 
 $t_{\rm 56Co}=113.5$ days, respectively. $\tau$ is given by 
 \begin{equation}
 \tau = 1000(Mej^{2}/M_{\odot})(E_{k}^{-1}/10^{51} {\rm erg~s}^{-1})(t/day)^{-2},
 \end{equation}
 where $M_{ej}$ is the mass and $E_{k}$ is the 
 kinetic energy of the ejected material. 
 
 The ratio of the 
 ejected mass to the kinetic energy is given by 
 ($M_{ej}/M_{\odot}$)$^{2}$ $\cdot$ 
 ($E_{k}$/10$^{51}$ erg~s$^{-1}$)$^{-1}=2.45$ by comparing the analytical light
 curve with our observed light curve. This degeneracy can be 
 resolved by adding photospheric velocity data. 
 As a proxy of the photospheric velocity, we used the 
 Si~{\sc ii} line velocity, which was $\sim$7000km~s$^{-1}$ 
 (see \S 3.3). Note that the Si~{\sc ii} velocity was 
 slightly larger than the photospheric velocity \citep{Fisher1997}.
 This should result in a kinetic energy increase
 or mass decrease. 
 We calculated the ejected mass to be $M_{ej}\sim1.15$~M$_{\odot}$ and 
 the kinetic energy to be $E_{K}\sim0.5$ $\times$10$^{51}$ 
 erg~s$^{-1}$  , while \citet{Stritzinger2015} 
 estimated the ejected mass as $1.4-1.9$ M$_{\odot}$ using their 
 calculations.
 We emphasize that our analysis contains 
 some uncertainties. 
 For example, the same opacity was assumed as that of a normal SN Ia.
   The effective opacity may have a strong dependence 
 on the temperatures \citep[e.g., ][]{Hoeflich1992}, i.e., 
 more luminous events can have higher opacities.
 It is, in fact, one of the promising explanation for
 the width-luminosity relation of normal SNe Ia
 \citep{Hoeflich1996,Nugent1997,Maeda2003,Kasen2007b,Baron2012}.
 If this trend is also true for SNe Iax,
 the effective opacity in the ejecta of SN 2012Z may be lower
 than in the ejecta of normal SNe Ia.
 Therefore, our estimated mass could be just a lower limit.
 Furthermore, a quantitative uncertainty exists in the density profile. 
 Thus, detailed 
 radiative transfer calculations are desirable. 
 In \S 4.3, we use only $E_{k}$ and $M_{ej}$ to
 estimate the upper limit of the shock luminosity.
 The late-phase luminosity could not be 
 explained completely using our analytical light curves. The diversity of 
 the decline rates may be significant among SNe Iax during the 
 late phase \citep{Stritzinger2015}.


\section{Discussion}
 \subsection{Comparison of the NIR properties with other SNe Iax}

 %
  We found no secondary maxima in the NIR light curves, and 
 the form of the color evolutions were similar to those of SNe 2005hk and 
 2008ha, as discussed in \S 3.1. Due to the small number of samples of 
 NIR light curves, the NIR absolute magnitudes have not yet been 
 well characterized for SNe Iax. 
 It remains ambiguous whether the rates of decline 
 correlate with the absolute magnitudes in the NIR bands.
 Such a correlation would mean that SN Iax explosions 
 may have a single origin \citep{Foley2013}.
 Here, we focus on the relationship between the NIR 
 absolute magnitudes and rates of decline of the light curves. 

\begin{figure*}
  \begin{center}
    \begin{tabular}{c}
      \resizebox{90mm}{!}{\includegraphics{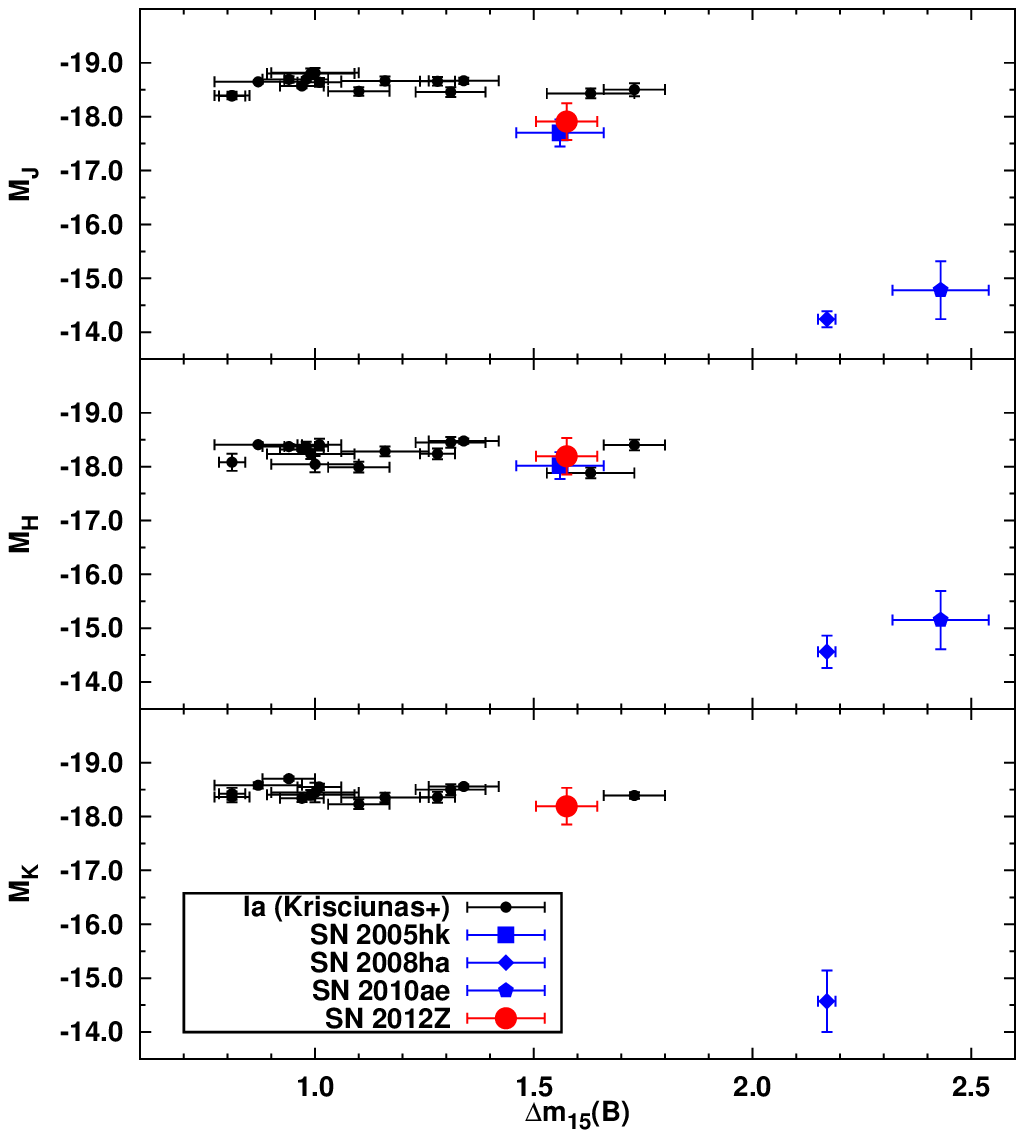}}  
    \end{tabular}
    \caption{$J$, $H$, and $K_{s}$-band absolute magnitudes plotted 
    as functions of the rate of decline $\Delta m_{15}$($B$) (red symbols). 
    Also shown are those of other 
    typical SNe Ia \citep{Krisciunas2004} (blue symbols). 
    The absolute magnitudes were calibrated 
    using distance moduli to the host galaxies.}
    \label{nirabs}
  \end{center}
\end{figure*}

 The absolute magnitudes and the rates of decline 
 of SN 2012Z in the $J$, $H$, and $K_{s}$ bands were compared 
 with those of other SNe Ia \citep{Krisciunas2004} and SNe Iax
 2005hk \citep{Phillips2007}, 2008ha \citep{Foley2010a}, and 
 2010ae \citep{Stritzinger2014a}, as shown in  Figure \ref{nirabs}. 
 In the $J$ and $H$ bands, the locations ($M_{J}\sim-18.1$ mag, 
 $M_{H}\sim-18.3$ mag, and $\Delta$ $m_{15}$($B$)$=1.6 \pm 0.1$ mag) 
 of SN 2012Z were close to those of 
 SN 2005hk. In contrast, the absolute magnitudes of SNe
 2008ha and 2010ae were significantly fainter than those of SN 2012Z. 
 The $K_{s}$-band absolute magnitude 
 ($M_{Ks}\sim$$-18.3$ mag) of SN 2012Z was
 similar to that of SNe Ia. Although the number of samples was 
 limited, we found a correlation between the 
 rates of decline 
 of the light curves and the absolute magnitudes.
 However, larger samples are required to conclude whether
 the origin of SNe Iax is homogeneous.
 

 \subsection{Comparison of rise times}
 
  We precisely calculated the rise time of a SN Iax 
 for the first time (see \S 3.2). 
 \cite{Phillips2007}
 reported a rise time of $15.0\pm0.1$ days for SN 2005hk 
 using a quadratic fit to the first two photometric 
 points. We used the same method to calculate the rise time of SN 2012Z. 
 
\begin{figure*}
  \begin{center}
    \begin{tabular}{c}
      \resizebox{90mm}{!}{\includegraphics{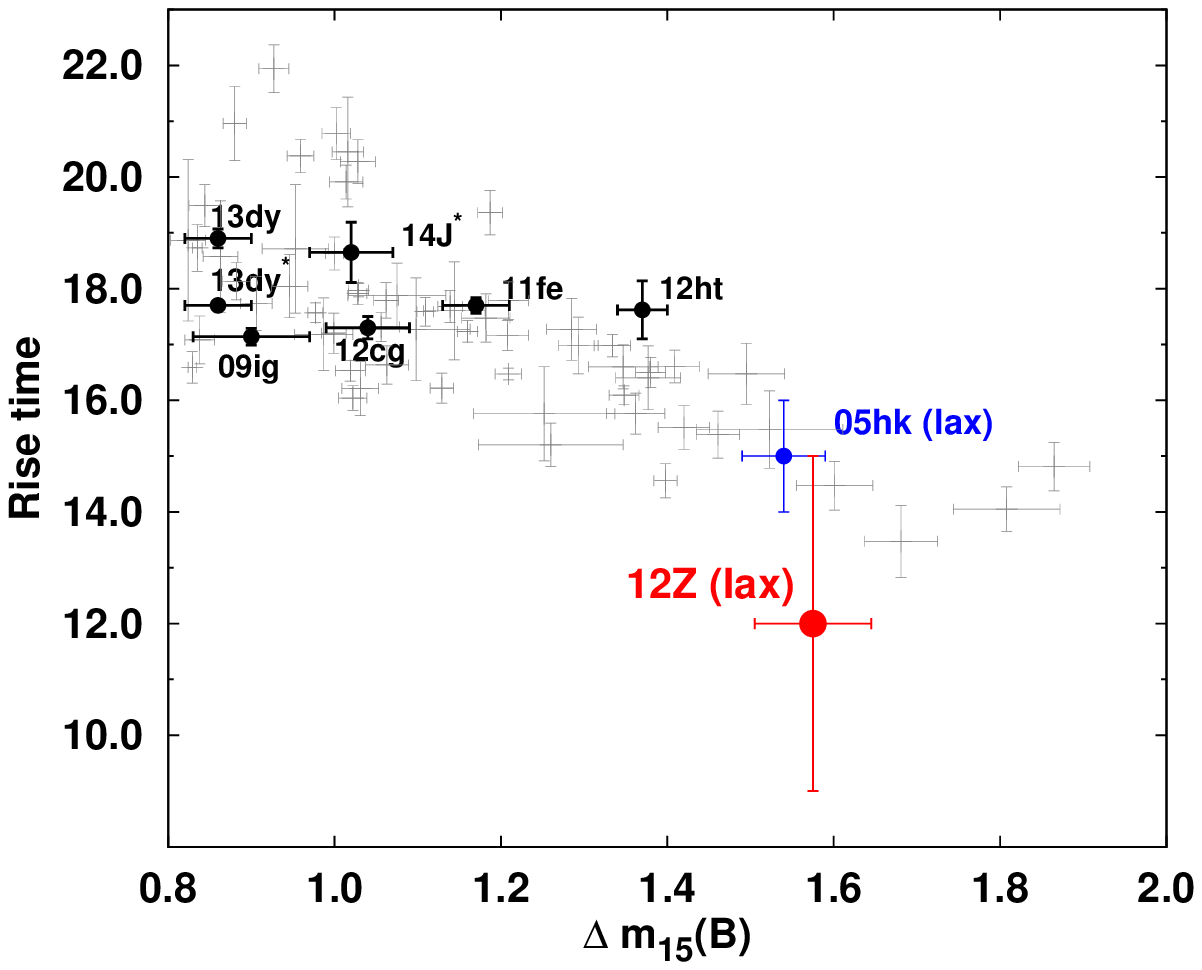}} 
    \end{tabular}
    \caption{Rate of decline $\Delta$m$_{15}$($B$) and rise time of SN 2012Z, 
  as well as those of SN 2005hk \citep{Phillips2007}, 
  2009ig \citep{Foley2012, Marion2013}, 
  2011fe \citep{Nugent2011, Pereira2013}, 2012cg \citep{Silverman2012a, Munari2013},
  2012ht \citep{Yamanaka2014}, 2013dy \citep{Zheng2013}, and 2014J \citep{Zheng2014}. 
  The rise time of SN 2013dy was found to be
  17.7 days based on a broken power-law fitting, whereas it was $\sim18.9$ 
  days using a quadratic fit \citep{Zheng2013}. 
  The gray cross symbols show template-fitting data from 
 \cite{Ganeshalingam2011}. 
 }
    \label{rtcmp}
  \end{center}
\end{figure*}

 Figure \ref{rtcmp} shows the rise time and $\Delta$m$_{15}$($B$) of these two 
 SNe Iax and other SNe Ia. \citet{Yamanaka2014} measured the 
 rise times of SNe Ia directly. 
 The rise times calculated using template-fitting methods are also 
 plotted. The rise times for SNe Ia were $17-19$ days, and the rise 
 time of SN 2012Z was  $12.0 \pm 3.0$ days, 
 which is significantly faster 
 than any other SNe Ia, including those calculated using 
 template-fitting extrapolation methods \citep{Ganeshalingam2011}. 

 The quadratic fit 
 of the light curves to SNe Iax has been problematic, 
 because of a paucity of data during the rising phase.
 The rise 
 times of individual SNe Iax have been discussed using various methods. 
 \cite{Foley2010a}
 calculated the rise time of SN 2008ha as $\sim$10 days using 
 a scaling method based on the light curves of SN 2005hk. 
 \cite{Narayan2011} also used a scaling method 
 to calculate the rise time of SN 2009ku, and reported 
 the rise time as $\sim18.2\pm3.0$ days. 
 \cite{McClelland2010} inferred the rise time of SN 2007qd
 as $10\pm2$ days using the previous upper limit magnitudes 
 prior to its discovery. 
 \cite{Foley2010b} reported that 
 the rise time of SN 2008ge should be between 
 9 and 27 days based on its spectral evolution. 

  These indicate that the rise times of SNe Ia exhibit a variety, 
 which may reflect the diversity of the 
 $^{56}$Ni distributions in the outer layer 
 \citep{Piro2012,Piro2013,Piro2014}. 
 The short rise time of SN 2012Z suggests that the mixing of the ejecta
 produces the extended
 $^{56}$Ni distributions \citep{Piro2014}. 
 Furthermore, the NIR light curves exhibit no secondary 
 maxima (see \S 3.1). This indicates that ejecta 
 mixing likely occurred \citep{Blinnikov2006,Kasen2006}. 
 The homogeniety of the NIR light curves 
 implies that mixing is a common property shared by SNe Iax.
 Thus, the theoretical models must 
 simultaneously describe the homogeneous  
 NIR light curves and the diverse rise times. 

    It is noted that the opacity can also affect 
 the rise time. Lower effective opacity results in the faster rise of 
 the light curves. As mentioned in \S 3.5. a lower luminosity may 
 give a lower opacity of the ejecta, and the effective opacity for 
 SN 2012Z may be lower than that in normal SNe Ia \citep{Stritzinger2015}.

  The variety of rise times may result in an over- or under-
 estimate of the $^{56}$Ni masses. To determine 
 the rise time with improved accuracy, 
 a high-cadence survey of the SN discoveries 
 at nearby galaxies is required.

\subsection{Progenitor systems}


  In this section we discuss the nature of the possible 
 companion star.
 The rising curves were satisfactorily fitted using a quadratic 
 function (see \S 3.3). However, ejecta-companion 
 interaction models predict 
 bright shock luminosity during the rising phase, especially for the 
 bluer bands \citep{Kasen2010}, yet such signatures were not 
 observed in our measurements. The existence of a giant 
 companion is, therefore, not supported; however, we note that the strength
 of this signal is expected to be dependent on the 
 viewing direction.

 The environments around SNe Iax are similar 
 to those of core collapse SNe IIP \citep{Lyman2013}. 
 For SN 2012Z, the age of the progenitor has been estimated to 
 be $30-50$ Myr \citep{Lyman2013}. Recently binary
 evolution studies \citep{XMeng2014,BWang2014,Liu2015} 
 have shown that the delay times for explosions 
 are as less than 100 Myr for 
 WD + He binary systems. We found that the H~{\sc ii} 
 region existed at the projected distance of $\sim90$pc to 
 SN 2012Z, indicating a young population as also 
 inferred by the low metallicity.
 This is consistent with the 
 compact blue source in the pre-explosion 
 image \citep{McCully2014}.

\begin{figure*}
  \begin{center}
    \begin{tabular}{c}
      \resizebox{160mm}{!}{\includegraphics{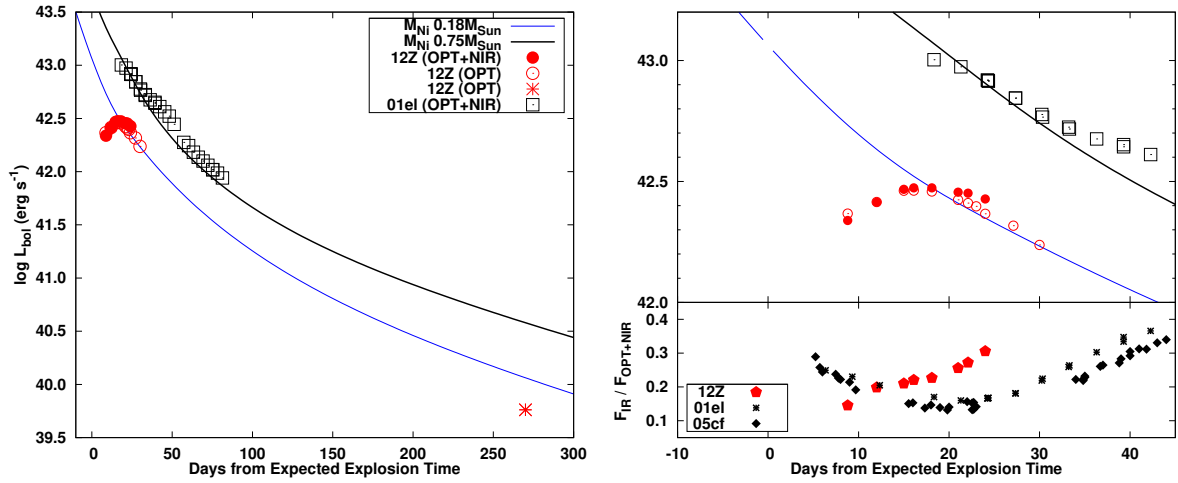}} 
    \end{tabular}
    \caption{(Left panel) Quasi-bolometric light curves of SN 2012Z 
    and normal SN Ia 2001el, both at the estimated 
    explosion date and 300 days later. The open circles 
    show the quasi-bolometric light curve integrated using 
    $BVRI$-band data, and the filled circles show the light curves integrated using 
    $BVRIJHK_{s}$-band data. 
    The late-phase luminosity at $t=280$ was the only $BVRI$-band 
    integration point.
    The plotted $BVRI$-band curves were corrected for by 
    multiplying the original flux by a factor of 1.25.
    For SN 2001el, the $BVRIJHK_{s}$-band data were integrated, as 
    shown by the black open-square symbols.
    The blue curve shows the analytical
    result for $M(^{56}{\rm Ni})=0.18 M_{\odot}$, $M_{\rm ej}=1.15 M_{\odot}$, 
    and $E_{\rm K}=0.5$ erg~s$^{-1}$. The black curve shows that with
    $M(^{56}{\rm Ni})=0.75 M_{\odot}$, $M_{\rm ej}=1.4 M_{\odot}$ and 
    $E_{\rm K}=1.4 \times 10^{51}$ erg~s$^{-1}$.
    (Right top panel) Same as the left panel, showing a close-up quasi-bolometric 
    light curve of the expected explosion date and 50 days later. 
    (Right bottom panel) Evolution of the ratio 
    of the integrated $JHK_{s}$-band to the $BVRIJHK_{s}$-band flux, including 
    data for SNe Ia 2001el and 2005cf \citep{XWang2009}.} 
    \label{qbl}
  \end{center}
\end{figure*}

 The upper limit of the progenitor radius was calculated following 
 \cite{Yamanaka2014}. Shock breakout is expected when 
 the early phase when the ejecta bursts through a progenitor, 
 and the deposited energy is released during the early phase
 \citep{Rabinak2011,Piro2013}. This emission 
 becomes more luminous when the progenitor has a larger 
 radius. In the extreme case, the upper limit of 
 luminosity for SN 2011fe 
 was given on just 0.2 days after the explosion 
 \citep{Bloom2012}. 
 Its progenitor radius has been reported as limited 
 to at most 0.02$R_{\odot}$ \citep{Bloom2012}. For SN 2012Z, 
 the luminosity was obtained at 2.0 days following the explosion 
 \citep{Cenko2012b}. 
 To eliminate uncertainties in the models, we 
 scaled the upper limit of the luminosity of SN 2011fe \citep{Bloom2012} 
 to that of SN 2012Z \citep[see ][ for details]{Yamanaka2014}. 
 We assumed the kinetic energy was 
 $E_{K}=0.5\times10^{51} erg~s^{-1}$ and the mass of 
 the ejecta was $M_{ej}=1.15 M_{\odot}$ 
 (see \S 3.5 and Figure \ref{qbl}).
 Note that the normalized density profile of the progenitor 
 as well as the opacity were assumed to be the same as those of SN 2011fe.
 We obtained an upper limit for the radius of SN 2012Z as $\sim$2.0 $R_{\odot}$.
 This ruled out an extended progenitor star like a 
 red giant \citep{Rabinak2011}.
 Scenarios with WD progenitors (both single degenerate and
 double degenerate) were therefore supperted.

 \subsection{Ejecta Structure}

  The emission lines during the late epoch were 6--8 times 
 broader than those of SN 2005hk, whereas the early-phase properties were
 similar.
 The artificially broadened line profiles of 
 SN 2005hk matched with those of SN 2012Z, as shown in Figure \ref{spcl}
 \citep{Sahu2008,Stritzinger2015}. 
 This suggests that the temperature and density of inner ejecta
 of SN 2012Z were similar to those of SN 2005hk; however,
 the velocity and thus spatial extent of the emission region
 were larger for SN 2012Z. 
 In contrast to the large difference in the late-phase spectra, the
 early-phase spectra and the light curves of SN 2005hk and SN 2012Z
 were similar.
 Thus, we speculate that the ejecta of SN 2005hk and SN 2012Z had
 a similar structure in the outer layer, whereas 
 the ejecta of SN 2012Z had a more extended inner core.  



 \subsection{Explosion models}

  Here, we compare the observed properties of
 SN 2012Z with the characteristic features predicted 
 by the following explosion models: 
 pure deflagration, failed deflagration, 
 double detonation, pulsating delayed detonation, 
 and core collapse. Table 7 lists a summary of these comparisons.

  A pure deflagration model \citep{Gamezo2003} has been 
 suggested for SNe Iax \citep{Phillips2007}. 
 The light curves and spectral evolutions have been 
 synthesized via radiative transfer calculations 
 \citep[][]{Blinnikov2006,Ropke2007,Long2014} based on 
 three-dimensional deflagrations models. 
 The small amount of ejected $^{56}$Ni and 
 the slow expansion 
 are consistent with those of SNe Iax. 
 There was also no secondary maximum in the theoretical NIR 
 light curves, which is interpreted as a mixing within 
 the ejecta \citep{Kasen2006}; this is 
 consistent with the observations. 
 However, the rise time predicted by the model was
 significantly longer than our calculated rise time \citep{Blinnikov2006}. 
 The luminosity of SNe 2008ha and 2010ae may 
 be too faint to compare to the predictions 
 from such explosion models.

  A failed deflagration model has also been suggested for 
 SNe Iax \citep{Jordan2012,Kromer2013,Fink2014}.
  \cite{Fink2014} indicated that part of the ejecta is 
 bound due to the small input energy. The mass of the compact remnant 
 has been predicted to be $\sim1.0~M_{\odot}$.
  The small ejected mass is qualitatively consistent with our estimate
 of the ejecta mass of SN 2012Z, which is potentially smaller than
 the Chandrasekhar limiting mass. The predicted mass in this particular
 model is smaller than our estimate, but we note there are uncertainties
 both in models and observationally derived ejecta mass.
 \cite{Fink2014} constructed synthetic spectra using 
 their models; the blue continuum, highly excited features, and 
 low velocities of the elements in the spectra were similar to 
 the spectra of SNe 2005hk and 2012Z.
 The form of the light curve and the absolute luminosities were also 
 well reproduced \citep{Fink2014}.  
 \cite{Kromer2013} reported that the 
 rise time in their models was $\sim$11 days, which is 
 consistent with that of SN 2012Z. A 
 compact remnant may also be consistent with the red source of 
 SN 2008ha detected 1500 days after the explosion 
 using HST archival images \citep{Foley2014}.


  Detection of possible He donors in the pre-explosion 
 image \citep{McCully2014} may indicate 
 a double-detonation scenario, triggered by 
 He accretion due to a companion star \citep{Nomoto1982b,Woosley1994}. 
 Although 
 our spectra do not exhibit any He~{\sc i} emission lines, 
 a few SNe Iax do \citep{Foley2013}. 
 However, the expected rise time, expansion 
 velocity, and absolute luminosity are 
 quite different from our data \citep{Sim2013}.
 This suggests that the detected pre-SN point source was 
 not a He donor, or challenges a theoretical model that an He 
 accretion results in the double-detonation.

 \citet{Stritzinger2015} compared their observations
 of SN 2012Z with the pulsational delayed detonation (PDD) 
 model, in which a detonation is triggered following a 
 failed deflagration \citep{Bravo2006}.
 \citet{Stritzinger2015} reported that the decline of the 
 line velocity in Ca~{\sc ii} H\&K and the flat-topped line 
 profile of [Fe~{\sc ii}] $\lambda$1.64 $\mu$m during the late phase 
 supports the PDD model. 
 The rapid rise and the low luminosity can also be reproduced 
 by the PDD model \citep{Baron2012}.  The higher
 luminosity should follow the later maximum in the NIR bands. 
 Such effect may explain the single peak in the $I$, 
 $J$, $H$, and $K_{s}$ bands \citep{Hoeflich1995}.
 The PDD may be another attractive scenarios in these aspects.

  Finally, we compare our observations with the behavior predicted by 
 a core collapse scenario. \cite{Moriya2010}  
 reported that the wide range of absolute luminosities 
 observed in SNe Iax
 could be reproduced by this model. The rapid rise may also be 
 explained; however, the quasi-bolometric light curves 
 and the Si~{\sc ii} velocity favor $M_{ej}=1.15 M_{\odot}$ 
 and $E_{k}=0.5\times10^{51}$ erg~s$^{-1}$, which cannot be 
 reproduced by a core collapse scenario.

 Table 7 lists a summary of these comparisons. 
 The failed deflagration model is the most favored scenario to describe
 the observed properties of SN 2012Z, as well as other SNe Iax. 
 However, as discussed in \S 4.2, SNe Iax exhibit varied 
 rise times; 
 it follows that the $^{56}$Ni distributions in the outer part of the ejecta 
 may also be diverse among SNe Iax. 
 More comprehensive explosion models are required to 
 describe both the diversity and homogeneity.

   \section{Conclusions}

  We have carried out optical and near-infrared observations of the SN Iax 2012Z 
 from immediately after the explosion to $t=+261$ d. We have described long-duration
 observations of the $J$, $H$, and $K_{s}$-band light curves of a SN Iax 
 for the first time. Our findings are summarized as follows.

  \begin{itemize}
    \item There were no secondary maxima in the $J$, $H$, and $K_{s}$ bands. The 
    NIR light curves and color evolutions of SN 2012Z were very similar 
    to those of other SNe Iax. 
    \item The $J$ and $H$-band absolute magnitudes were $M_{J}\sim-18.1$ mag and 
    $M_{H}\sim$$-18.3$ mag, and the rate of 
    decline of the $B$-band light curve was 
    $\Delta$ $m_{15}$($B$)$=1.6 \pm 0.1$ mag. These are very similar to those of the 
    bright SN Iax 2005hk; however, SN 2012Z was more luminous than the
    faint SNe Iax 2008ha and 2010ae. 
    \item The rise time was calculated to be $12.0 \pm 3.0$
    days, 
    which is significantly shorter than $15.0\pm0.1$ days for SN 2005hk. 
    The rise times of SNe Iax may exhibit significant diversity. 
    \item The explosion site was characterized by low metallicity and a 
    moderate star-forming environment; this is consistent with the 
    binary evolution models of an He star. However, the popular explosion 
    scenario for such a system (i.e., double detonation) is not compatible 
    to the observed SN properties.
   \end{itemize}
  The rapid rise indicates that $^{56}$Ni was present in the outer 
 layer. The single peak of the NIR light curves also suggests mixing 
 of iron-peak elements throughout the ejecta.  
 These properties may be also explained by the opacity effect.
 The rise indicates that the progenitor was 
 more compact than a red giant.

   Comparing our observations with the theoretical models,
 the most probable scenario appears to be the failed deflagration model,
 although PDD model \citep{Stritzinger2015} is not excluded.
 The large diversity of the rise times observed in SNe
 Iax requires a more comprehensive theoretical analysis.
 The observations indicate that they 
   may originate from a homogeneous population.
 There appears to be a diverse range of rise times, however, 
 and bright SNe Iax samples are required to further investigate this.
   A high-cadence survey of nearby galaxies is therefore encouraged. 

 \acknowledgements
 This work was partly supported by the Grant-in-Aid for 
 Scientific Research from JSPS (26800100) and the 
 WPI Initiative, MEXT.
 This work was also supported by the Optical and 
 Near-infrared Astronomy Inter-University Cooperation Program, and
 the Hirao Taro Foundation of the Konan University Association for
 Academic Research.


  \newpage
  \clearpage
  \newpage

\begin{deluxetable}{lll}
  \tabletypesize{\scriptsize}
    \tablecaption{Properties of SN 2012Z and its host galaxy NGC 1309.} 
  \tablewidth{0pt}
   \startdata
    \hline \hline
     {\bf SN 2012Z}              &                                   &  References \\
     \hline 
     $\alpha$ (2000.0)          & 03h22m05.35s                      & 1 \\
     $\delta$ (2000.0)          & $-$15$^{\circ}$23$'$15$''$            & 1 \\
     Discovered magnitude       & 17.6 (Unfilter)                   & 1 \\
     Discovery date (UT)        & 2012 Jan 29.150                  & 1 \\
     Discovery date (MJD)       & 55955.15                          & 1 \\
     Discoverer                 & Lick Observatory Supernova Search & 1 \\
     SN Type                    & Ia-pec (Type Iax)                 & 2,3 \\ 
     Galactic extinction        & $E$($B-V$)$=0.036$                  & 4 \\
     host galactic extinction  & $E$($B-V$)$=0.0$                    & 5 \\             
     $B$-band maximum date      & 55965.8                           & 5 \\
     $B$-band maximum magnitude & 14.74 mag                            & 5 \\
     absolude magnitude (M$_B$) & $-17.61$ mag                      & 5 \\
     $\Delta m_{15}$($B$)     & $1.57 \pm 0.07$ mag               & 5 \\
     $v_{max}$(Si~{\sc ii})     & $6980 \pm 80$ km~s$^{-1}$        & 5 \\ 
     Explosion date (MJD)       & $55953.4 \pm 0.2$                 & 5 \\
     Rise time                  &  $12.0 \pm 3.0$ day                & 5 \\
     $^{56}$Ni mass             & $\sim$0.18 $M_{\odot}$             & 5 \\
     \hline
     \hline
     {\bf NGC 1309}             &                                   &            \\
     \hline
     $\alpha$ (2000.0)          & 03h22m06.5s                       & 6 \\
     $\delta$ (2000.0)          & $-$15$^{\circ}$ 24$'$00$''$       & 6 \\
     morphology type            & SA(s)bc                           & 6 \\
     redshift                   & 0.007                             & 6 \\
     distance                   & $29.8 \pm 3.8 $ Mpc           & 6 \\
     distance modulus           & $32.4 \pm 0.3$ mag               & 6 \\ 
     apparent magnitude ($B$)     & 11.7 mag                        & 6 \\
     absolute magnitude (M$_{B}$) & -20.65 mag                      & 6 \\
     metallicity                & 12+log(O/H)$<8.51 \pm 0.31$       & 5 
   \enddata
\tablenotetext{1}{\cite{Cenko2012a}}{}
\tablenotetext{2}{\cite{Cenko2012c}}{}
\tablenotetext{3}{\cite{Meyer2012}}{}
\tablenotetext{4}{\cite{Schlafly2011}}{}
\tablenotetext{5}{This paper}{}
\tablenotetext{6}{$NED$}{}
\end{deluxetable}


\begin{deluxetable}{lllll}
  \tabletypesize{\scriptsize}
    \tablecaption{Summary of the properties of the telescopes, instruments, and observatories.} 
  \tablewidth{0pt}
   \startdata
    \hline
    \hline
     Observatory     &  Telescope           &  Instruments & Filters/Resolutions & Number of nights \\
    \hline
     NO$^{a}$        &  1.6m Pirka          &  MSI$^{b}$      & $U$, $B$, $V$, $R$, $I$  & 4     \\
     AO$^{c}$        &  0.5m MITSuME$^{d}$  &  CCD            & $g'$, $R$, $I$           & 10    \\
     KAO$^{e}$       &  1.3m Araki          &  LOSA/F2$^{f}$  & R=600                    & 4      \\
     OAO$^{g}$       & 1.88m                &  ISLE$^{h}$     & $J$, $H$, $K_{s}$    & 1     \\
     OAO             & 0.91m                &  WFC$^{i}$      & $J$, $H$, $K_{s}$    & 5      \\
     OAO             & 0.5m MITSuME$^{d}$   &  CCD            & $g'$, $R$, $I$           & 14      \\
     NHAO$^{j}$      & 2.0m Nayuta          &  NIC$^{k}$      & $J$, $H$, $K_{s}$       & 9      \\
     HHO$^{l}$       & 1.5m Kanata          &  HOWPol$^{m}$   & $B$, $V$, $R$, $I$       & 23      \\
     HHO             & 1.5m Kanata          &  HOWPol         & R=400                    & 12      \\
     IO$^{n}$        & 1.0m                 &  IR Cam         & $J$, $H$, $K_{s}$    & 12       \\
     IAO$^{o}$       & 1.0m MITSuME$^{d}$   & CCD             & $g'$, $R$, $I$           & 7        \\
     SAAO$^{p}$      & 1.4m IRSF            & SIRIUS$^{q}$    & $J$, $H$, $K_{s}$    & 20      \\
     KO              & 1.88m                & CCD             & $B$, $V$, $R$, $I$       & 2      \\
     NAOJ$^{r}$      & 8.2m Subaru          & FOCAS$^{s}$     & $B$, $V$, $R$, $I$       & 1      \\
     NAOJ            & 8.2m Subaru          & FOCAS           & R=650                    & 1      
   \enddata
\tablenotetext{a}{Nayoro Observatory}{}
\tablenotetext{b}{Multispectral Imager \citep{Watanabe2012}}{}
\tablenotetext{c}{Akeno Observatory}{}
\tablenotetext{d}{Multicolor Imaging Telescopes for Survey and Monstrous 
 Explosions \citep{Kotani2005}}{}
\tablenotetext{e}{Koyama Astronomical Obsrvatory}{}
\tablenotetext{f}{\cite{Shinnaka2013}}{}
\tablenotetext{g}{Okayama Astrophysical Observatory}{}
\tablenotetext{h}{InfraRed Imager/Spectrograph \citep{Yanagisawa2006}}{}
\tablenotetext{i}{Wide Field Camera \citep{Yanagisawa2014}}{}
\tablenotetext{j}{Nishi-Harima Astronomical Observatory}{}
\tablenotetext{k}{Nishi-harima Infrared Camera}{}
\tablenotetext{l}{Higashi-Hiroshima Observatory}{}
\tablenotetext{m}{Hiroshima One-shot Wide-field Polarimeter \citep{Kawabata2008}}{}
\tablenotetext{n}{Iriki Observatory}{}
\tablenotetext{o}{Ishigaki-jima Astronomical Observatory}{}
\tablenotetext{p}{South African Astronomical Observatory}{}
\tablenotetext{q}{Near-infrared simultaneous three-band camera \citep{Nagayama2003}}{}
\tablenotetext{r}{National Astronomical Observatory of Japan}{}
\tablenotetext{s}{Faint Object Camera and Spectrograph for Subaru Telescope \citep{Kashikawa2002}}{}
\end{deluxetable}





\begin{deluxetable}{cccccc}
  \tabletypesize{\tiny}
   \tablecaption{Log of the spectroscopic observations.} 
  \tablewidth{0pt}
   \startdata
    \hline \hline
    \multicolumn{1}{c}{Date} & MJD  & Phase & Covarage & Resolution & Instruments  \\
     \hline
2012 Feb 3 & 55960.4 & -5.0 & 4000-8000 \AA & 10.9 \AA & LOSA/F2 \\ 
2012 Feb 4 & 55961.5 & -3.9 & 4500-9000 \AA & 15.0 \AA & HOWPol \\  
2012 Feb 8 & 55965.4 & 0.0 & 4000-8000 \AA & 10.9 \AA & LOSA/F2 \\  
2012 Feb 9 & 55966.5 & 1.1 & 4500-9000 \AA & 15.0 \AA & HOWPol \\   
2012 Feb 11 & 55968.4 & 3.0 & 4500-9000 \AA & 15.0 \AA & HOWPol \\   
2012 Feb 11 & 55968.5 & 3.1 & 4000-8000 \AA & 10.9 \AA & LOSA/F2 \\  
2012 Feb 12 & 55969.5 & 4.1 & 4500-9000 \AA & 15.0 \AA & HOWPol \\   
2012 Feb 18  & 55975.5 & 10.1 & 4500-9000 \AA & 15.0 \AA & HOWPol \\  
2012 Feb 19 & 55976.4 & 11.0 & 4500-9000 \AA & 15.0 \AA & HOWPol \\  
2012 Feb 19 & 55976.4 & 11.0 & 4000-8000 \AA & 10.9 \AA & LOSA/F2 \\ 
2012 Feb 20 & 55977.5 & 12.1 & 4500-9000 \AA & 15.0 \AA & HOWPol \\
2012 Feb 26 & 55983.5 & 18.1 & 4500-9000 \AA & 15.0 \AA & HOWPol \\
2012 Feb 27 & 55984.5 & 19.1 & 4500-9000 \AA & 15.0 \AA & HOWPol \\
2012 Mar 11 & 55997.5 & 32.1 & 4500-9000 \AA & 15.0 \AA & HOWPol \\
2012 Mar 13 & 55999.4 & 34.0 & 4500-9000 \AA & 15.0 \AA & HOWPol \\
2012 Mar 15 & 56001.7 & 36.3 & 4500-9000 \AA & 15.0 \AA & HOWPol \\
2012 Oct 23 & 56226.7 & 261.3 & 4600-9600 \AA & 9.1 \AA & FOCAS
   \enddata
\end{deluxetable}

\begin{deluxetable}{ccccc}
  \tabletypesize{\tiny}
   \tablecaption{Near-infrared photometry data of SN 2012Z.} 
  \tablewidth{0pt}
   \startdata
    \hline \hline
    \multicolumn{1}{c}{MJD} 
   & $J$ & $H$  & $K_{\rm S}$ & Instruments  \\
     \hline
55960.4 & 15.006(0.109) & 15.334(0.270) & 15.081(0.485) & NIC \\
55960.4 & 15.610(0.210) & $\cdots$ & $\cdots$ & WFC \\
55961.4 & 14.980(0.224) & 15.079(0.217) & 14.684(0.311) & NIC \\
55961.4 & 15.010(0.020) & 15.210(0.030) & 15.140(0.050) & ISLE \\
55961.5 & 14.931(0.037) & $\cdots$           & $\cdots$           & IR Cam \\  
55961.5 & 15.060(0.080) & 15.270(0.180) & 15.130(0.280) & WFC \\
55961.8 & 14.980(0.040) & 15.280(0.050) & 15.160(0.100) & IRSF \\
55962.8 & 14.910(0.050) & 15.030(0.050) & 15.040(0.100) & IRSF \\
55964.8 & 14.620(0.050) & 14.890(0.060) & 14.910(0.160) & IRSF \\
55965.8 & 14.620(0.030) & 14.740(0.030) & 14.580(0.060) & IRSF \\
55966.4 & $\cdots$      & 14.602(0.064) & 14.419(0.156) & NIC \\
55966.5 & 14.659(0.030) & $\cdots$           & $\cdots$           & IR Cam \\   
55966.8 & 14.550(0.020) & 14.630(0.020) & 14.620(0.060) & IRSF \\
55967.8 & 14.520(0.020) & 14.610(0.020) & 14.530(0.050) & IRSF \\
55968.4 & 14.411(0.031) & 14.440(0.051) & 14.334(0.112) & NICS \\
55968.4 & 14.520(0.050) & $\cdots$      & 14.250(0.090) & WFC \\
55968.5 & 14.387(0.037) & $\cdots$           & $\cdots$            & IR Cam \\    
55968.8 & 14.430(0.020) & 14.470(0.020) & 14.400(0.040) & IRSF \\
55969.4 & 14.486(0.010) & 14.397(0.007) & 14.257(0.016) & NIC \\
55969.5 & 14.550(0.060) & $\cdots$ & 14.440(0.110) & WFC \\
55969.7 & 14.430(0.020) & 14.410(0.020) & 14.310(0.040) & IRSF \\
55970.7 & 14.430(0.020) & 14.360(0.020) & 14.270(0.040) & IRSF \\
55971.8 & 14.410(0.020) & 14.320(0.020) & 14.290(0.040) & IRSF \\
55972.5 & 14.434(0.031) & $\cdots$           & $\cdots$           & IR Cam \\
55973.5 & 14.348(0.033) & 14.262(0.061) & $\cdots$           & IR Cam \\
55973.8 & 14.430(0.020) & 14.250(0.020) & 14.120(0.040) & IRSF \\
55974.4 & 14.255(0.006) & 14.255(0.006) & 14.062(0.060) & NIC \\
55974.8 & 14.460(0.020) & 14.280(0.020) & 14.160(0.040) & IRSF \\
55975.8 & 14.340(0.020) & 14.170(0.020) & 14.130(0.040) & IRSF \\
55976.4 & 14.382(0.055) & 14.094(0.072) & 13.753(0.203) & NIC \\
55977.4 & 14.333(0.032) & 14.055(0.014) & 13.917(0.034) & NIC \\
55977.4 & 14.440(0.050) & 14.230(0.070) & 13.940(0.080) & WFC \\
55977.5 & 14.345(0.034) & $\cdots$           & $\cdots$           & IR Cam \\
55977.8 & 14.400(0.020) & 14.070(0.020) & 14.040(0.040) & IRSF \\
55983.5 & 14.280(0.033) & $\cdots$           & $\cdots$           & IR Cam \\  
55990.8 & 14.380(0.030) & 14.080(0.020) & 14.200(0.050) & IRSF \\
55993.7 & 14.380(0.030) & 14.170(0.020) & 14.300(0.060) & IRSF \\
55995.4 & $\cdots$           & 14.536(0.254) & $\cdots$           & IR Cam \\  
55996.4 & 14.705(0.030) & $\cdots$           & $\cdots$           & IR Cam \\ 
55996.8 & 14.550(0.030) & 14.290(0.030) & 14.320(0.060) & IRSF \\
55999.4 & 14.705(0.030) & $\cdots$           & $\cdots$           & IR Cam \\ 
56000.4 & 14.653(0.205) & $\cdots$           & $\cdots$           & IR Cam \\  
56006.8 & 15.510(0.090) & 14.730(0.060) & 14.600(0.150) & IRSF \\
56021.7 & 15.960(0.120) & 15.380(0.080) & 16.580(0.720) & IRSF \\
56027.7 & 16.300(0.150) & 15.250(0.080) & 16.420(0.520) & IRSF
   \enddata
\end{deluxetable}


\begin{deluxetable}{cccccccc}
  \tabletypesize{\scriptsize}
    \tablecaption{Standard magnitudes of the comparison stars in the local field of SN 2012Z.} 
  \tablewidth{0pt}
   \startdata
    \hline \hline
ID  & $U$$^{a}$              & $B$                    & $g'$              & $V$                    & $R$                    & $I$                    & $z'+Y$$^{b}$ \\
     \hline
$1$ & $13.53$ $\pm$ $0.31$ & $13.50$ $\pm$ $0.05$ & $13.03 \pm 0.02$  & $12.77$ $\pm$ $0.05$ & $12.35$ $\pm$ $0.03$ & $11.94$ $\pm$ $0.03$ & $12.31$ $\pm$ $0.01$ \\
$2$ & $13.45$ $\pm$ $0.73$ & $13.42$ $\pm$ $0.05$ & $12.85 \pm 0.02$  & $12.39$ $\pm$ $0.05$ & $11.77$ $\pm$ $0.03$ & $11.26$ $\pm$ $0.03$ & $11.56$ $\pm$ $0.01$ \\
$3$ & $13.35$ $\pm$ $0.18$ & $13.32$ $\pm$ $0.05$ & $13.00 \pm 0.02$  & $12.68$ $\pm$ $0.05$ & $12.35$ $\pm$ $0.03$ & $12.07$ $\pm$ $0.03$ & $\cdots$ \\
$4$ & $14.25$ $\pm$ $0.18$ & $14.22$ $\pm$ $0.04$ & $13.95 \pm 0.03$  & $13.58$ $\pm$ $0.04$ & $13.27$ $\pm$ $0.03$ & $12.90$ $\pm$ $0.03$ & $13.27$ $\pm$ $0.02$\\
$5$ & $\cdots$             & $16.46$ $\pm$ $0.05$ & $16.10 \pm 0.06$  & $15.89$ $\pm$ $0.04$ & $15.43$ $\pm$ $0.09$ & $15.07$ $\pm$ $0.07$ & $15.38$ $\pm$ $0.06$ \\
$6$ & $\cdots$             & $17.65$ $\pm$ $0.09$ & $\cdots$          & $17.06$ $\pm$ $0.06$ & $16.59$ $\pm$ $0.10$ & $16.22$ $\pm$ $0.07$ & $16.22$ $\pm$ $0.07$ \\
$7$ & $\cdots$             & $20.21$ $\pm$ $0.09$ & $\cdots$          & $18.88$ $\pm$ $0.06$ & $17.89$ $\pm$ $0.10$ & $17.06$ $\pm$ $0.07$ & $\cdots$ 
   \enddata
\tablenotetext{a}{The $U$-band magnitudes were converted using the $U-B$ and $B-V$ relations in the Landolt standard magnitudes sequence \citep{Landolt1992}. 
The $B-V$ color obtained via the photometric calibration with Kanata/HOWPol.}{} 
\tablenotetext{b}{The $z'+Y$-band magnitudes were calculated using the color relations between the $R$ and $z'+Y$ bands, which were obtained 
from photometric calibration using the Landolt standard stars SA95 96 \citep{Landolt1992,Smith2002}.}{}
\label{tb:ls}
\end{deluxetable}


\begin{deluxetable}{ccccccc}
  \tabletypesize{\scriptsize}
    \tablecaption{Parameters of the optical and near-infrared light curves of SN 2012Z.} 
  \tablewidth{0pt}
   \startdata
    \hline \hline
Filter & Maximum date (MJD) & Error & Maximum magnitude & Error & $\Delta m_{15}$($\lambda$) & Error \\
     \hline 
$B$ & 55965.8 & 3.0 & 14.738 & 0.001 & 1.58 & 0.07  \\
$V$ & 55971.8 & 0.20 & 14.470 & 0.008 & 0.75 & 0.05 \\
$R$ & 55974.2 & 0.04 & 14.209 & 0.002 & 0.52 & 0.01 \\
$I$ & 55977.5 & 0.12 & 13.986 & 0.001 & 0.44 & 0.003 \\
$J$ & 55982.0 & 0.19 & 14.310 & 0.008 & 0.27 & 0.02 \\
$H$ & 55983.9 & 0.04 & 14.027 & 0.002 & 0.32 & 0.01 \\
$K_{s}$ & 55981.3 & 0.23 & 14.027 & 0.005 & 0.33 & 0.02
   \enddata
\end{deluxetable}

\begin{deluxetable}{cccccc}
  \tabletypesize{\scriptsize}
    \tablecaption{Summary of whether the existing theoretical models are consistent with our observations.} 
  \tablewidth{0pt}
   \startdata
    \hline \hline
                    & $V_{ph}$ & Rise time & Spectra & LC  & Luminosity \\
     \hline
pure deflagration   & YES      & NO        & YES     & YES & YES   \\
failed deflagration & YES      & YES       & YES     & YES & YES   \\
PDD                 & NO       & YES       & ?       & YES  & YES  \\
double detonations  & NO       & NO        & NO      & NO  & NO   \\  
CC                  & YES      & YES       & ?       & NO  & YES
   \enddata
\end{deluxetable}

  \newpage
   \clearpage


\begin{deluxetable}{ccccccccc}
  \tabletypesize{\tiny}
\tablecaption{Optical photometry data of SN 2012Z.}
  \startdata
    \hline \hline
    \multicolumn{1}{c}{MJD}
    &  $U$ & $B$ & $g'$ & $V$ & $R$ & $I$ & $z'+Y$ & Instruments \\
 \hline 
55960.4 & $\cdots$ & $\cdots$ & 15.330(0.020) & $\cdots$ & 15.130(0.020) & 15.080(0.030) & $\cdots$ & MITSuME(O) \\
55960.5 & $\cdots$ & $\cdots$ & $\cdots$ & 15.273(0.230) & 15.096(0.046) & 15.054(0.052) & 15.364(0.023) & HOWPol \\
55961.4 & $\cdots$ & $\cdots$ & 15.160(0.030) & $\cdots$ & 14.930(0.020) & 14.850(0.030) & $\cdots$ & MITSuME(O) \\
55961.4 & $\cdots$ & $\cdots$ & $\cdots$ & $\cdots$ & 15.181(0.029) & 15.210(0.035) & $\cdots$ & MITSuME(A) \\
55961.5 & $\cdots$ & $\cdots$ & 15.125(0.004) & $\cdots$ & 14.973(0.004) & 15.050(0.010) & $\cdots$ & MITSuME(I) \\
55961.5 & $\cdots$ & 14.827(0.023) & $\cdots$ & 14.778(0.024) & 14.803(0.175) & 14.700(0.175) & 15.219(0.031) & HOWPol \\
55964.4 & $\cdots$ & $\cdots$ & 14.950(0.060) & $\cdots$ & 14.530(0.040) & 14.560(0.050) & $\cdots$ & MITSuME(O) \\
55964.5 & $\cdots$ & 14.735(0.026) & $\cdots$ & $\cdots$ & $\cdots$ & $\cdots$ & 14.828(0.030) & HOWPol \\
55965.4 & 14.072(0.106) & 14.922(0.012) & $\cdots$ & 14.871(0.050) & 14.273(0.077) & $\cdots$ & $\cdots$ & MSI \\
55965.4 & $\cdots$ & 14.832(0.049) & $\cdots$ & 14.661(0.048) & 14.495(0.033) & 14.425(0.027) & 14.736(0.039) & HOWPol \\
55965.4 & $\cdots$ & $\cdots$ & 14.738(0.047) & $\cdots$ & 14.423(0.100) & 14.279(0.073) & $\cdots$ & MITSuME(A) \\
55965.4 & $\cdots$ & $\cdots$ & 14.770(0.050) & $\cdots$ & 14.510(0.040) & 14.580(0.070) & $\cdots$ & MITSuME(O) \\
55966.4 & $\cdots$ & 14.741(0.008) & $\cdots$ & 14.598(0.003) & $\cdots$ & $\cdots$ & $\cdots$ & HOWPol \\
55966.4 & $\cdots$ & $\cdots$ & 14.714(0.047) & $\cdots$ & 14.385(0.117) & 14.277(0.078) & $\cdots$ & MITSuME(A) \\
55966.5 & $\cdots$ & $\cdots$ & 14.720(0.030) & $\cdots$ & 14.340(0.020) & 14.260(0.020) & $\cdots$ & MITSuME(O) \\
55966.5 & $\cdots$ & $\cdots$ & 14.790(0.030) & $\cdots$ & 14.360(0.020) & 14.270(0.030) & $\cdots$ & MITSuME(O) \\
55968.4 & $\cdots$ & 14.790(0.021) & $\cdots$ & 14.518(0.024) & 14.322(0.025) & 14.215(0.021) & 14.527(0.019) & HOWPol \\
55968.4 & $\cdots$ & $\cdots$ & 14.670(0.010) & $\cdots$ & 14.280(0.010) & 14.240(0.010) & $\cdots$ & MITSuME(O) \\
55969.5 & $\cdots$ & 14.824(0.022) & $\cdots$ & 14.506(0.042) & 14.284(0.041) & 14.192(0.026) & 14.504(0.018) & HOWPol \\
55969.6 & $\cdots$ & $\cdots$ & 14.669(0.005) & $\cdots$ & 14.300(0.005) & 14.404(0.009) & $\cdots$ & MITSuME(I) \\
55971.5 & $\cdots$ & 14.942(0.026) & $\cdots$ & 14.448(0.027) & 14.244(0.033) & 14.118(0.024) & 14.431(0.019) & HOWPol \\
55971.5 & $\cdots$ & $\cdots$ & 14.667(0.003) & $\cdots$ & 14.154(0.005) & 14.383(0.005) & $\cdots$ & MITSuME(I) \\
55972.5 & $\cdots$ & $\cdots$ & 14.820(0.050) & $\cdots$ & 14.150(0.020) & 14.040(0.030) & $\cdots$ & MITSuME(O) \\
55973.4 & $\cdots$ & $\cdots$ & 14.800(0.020) & $\cdots$ & 14.150(0.010) & 14.040(0.020) & $\cdots$ & MITSuME(O) \\
55974.4 & $\cdots$ & 15.264(0.037) & $\cdots$ & 14.500(0.030) & 14.204(0.031) & 14.008(0.024) & 14.323(0.024) & HOWPol \\
55974.4 & $\cdots$ & $\cdots$ & 14.784(0.063) & $\cdots$ & 14.159(0.101) & 13.998(0.078) & $\cdots$ & MITSuME(A) \\
55975.4 & $\cdots$ & $\cdots$ & 14.870(0.020) & $\cdots$ & 14.220(0.010) & 13.960(0.020) & $\cdots$ & MITSuME(O) \\
55975.4 & $\cdots$ & $\cdots$ & 14.893(0.059) & $\cdots$ & 14.217(0.107) & 13.925(0.063) & $\cdots$ & MITSuME(A) \\
55975.5 & $\cdots$ & 15.396(0.030) & $\cdots$ & 14.500(0.024) & 14.211(0.028) & 13.988(0.022) & 14.304(0.018) & HOWPol \\
55976.4 & $\cdots$ & 15.524(0.026) & $\cdots$ & 14.528(0.024) & 14.208(0.028) & 13.963(0.021) & 14.280(0.017) & HOWPol \\
55976.4 & $\cdots$ & $\cdots$ & 14.906(0.057) & $\cdots$ & 14.226(0.104) & 13.901(0.068) & $\cdots$ & MITSuME(A) \\
55976.4 & $\cdots$ & $\cdots$ & 15.000(0.020) & $\cdots$ & 14.220(0.010) & 13.980(0.010) & $\cdots$ & MITSuME(O) \\
55977.4 & $\cdots$ & 15.653(0.030) & $\cdots$ & 14.579(0.025) & 14.270(0.028) & 14.040(0.022) & 14.356(0.019) & HOWPol \\
55977.4 & $\cdots$ & $\cdots$ & 15.090(0.060) & $\cdots$ & 14.276(0.110) & 13.972(0.080) & $\cdots$ & MITSuME(A) \\
55977.4 & $\cdots$ & $\cdots$ & 15.100(0.020) & $\cdots$ & 14.200(0.010) & 13.980(0.010) & $\cdots$ & MITSuME(O) \\
55977.4 & $\cdots$ & $\cdots$ & $\cdots$ & 14.603(0.012) & $\cdots$ & $\cdots$ & $\cdots$ & MSI \\
55979.4 & $\cdots$ & 15.844(0.019) & $\cdots$ &  14.722(0.015) & $\cdots$ & $\cdots$ & $\cdots$ & MSI \\
55980.4 & $\cdots$ & $\cdots$ & 15.420(0.060) & $\cdots$ & 14.286(0.110) & 13.922(0.070) & $\cdots$ & MITSuME(A) \\
55980.5 & $\cdots$ & 16.171(0.382) & $\cdots$ & 14.740(0.057) & 14.288(0.079) & 13.933(0.058) & 14.253(0.077) & HOWPol \\
55983.4 & $\cdots$ & 16.690(0.039) & $\cdots$ & 14.944(0.025) & 14.422(0.027) & 14.048(0.022) & 14.369(0.066)& HOWPol \\
55983.4 & $\cdots$ & $\cdots$ & 15.670(0.050) & $\cdots$ & 14.410(0.020) & 14.010(0.020) & $\cdots$ & MITSuME(O) \\
55983.7 & $\cdots$ & 16.621(0.010) & $\cdots$ & 15.030(0.050) & 14.420(0.030) & 14.010(0.030) & $\cdots$ & Kottamia \\
55984.4 & $\cdots$ & 16.774(0.044) & $\cdots$ & 15.020(0.025) & 14.484(0.026) & $\cdots$ & $\cdots$ & HOWPol \\
55984.4 & $\cdots$ & $\cdots$ & 15.840(0.060) & $\cdots$ & 14.486(0.110) & 14.042(0.070) & $\cdots$ & MITSuME(A) \\
55985.4 & $\cdots$ & $\cdots$ & $\cdots$ & 15.192(0.034) & $\cdots$ & $\cdots$ & $\cdots$ & MSI \\
55985.7 & $\cdots$ & 16.909(0.030) & $\cdots$ & 15.334(0.009) & 14.578(0.009) & 14.153(0.013) & $\cdots$ & Kottamia \\
55986.5 & $\cdots$ & $\cdots$ & $\cdots$ & 15.234(0.041) & 14.574(0.036) & 14.168(0.026) &  $\cdots$& HOWPol \\
55987.4 & $\cdots$ & $\cdots$ & 16.170(0.170) & $\cdots$ & 14.626(0.110) & 14.212(0.080) & $\cdots$ & MITSuME(A) \\
55987.5 & $\cdots$ & $\cdots$ & 16.165(0.007) & $\cdots$ & 14.648(0.004) & 14.312(0.005) & $\cdots$ & MITSuME(I) \\
55988.5 & $\cdots$ & $\cdots$ & 16.195(0.034) & $\cdots$ & 14.701(0.011) & 14.272(0.020) & $\cdots$ & MITSuME(I) \\
55994.5 & $\cdots$ & $\cdots$ & 16.611(0.024) & $\cdots$ & 15.046(0.006) & 14.606(0.008) & $\cdots$ & MITSuME(I) \\
55996.4 & $\cdots$ & $\cdots$ & $\cdots$ & 15.883(0.044) & 15.172(0.037) & 14.648(0.028) & 14.974(0.035) & HOWPol \\
55997.4 & $\cdots$ & $\cdots$ & $\cdots$ & 15.758(0.034) & 15.202(0.039) & 14.695(0.026) & 15.020(0.021) & HOWPol \\
55998.4 & $\cdots$ & $\cdots$ & $\cdots$ & 15.807(0.038) & 15.243(0.032) & 14.714(0.024) & 15.040(0.021) & HOWPol \\
55999.5 & $\cdots$ & $\cdots$ & 17.000(0.190) & $\cdots$ & 15.220(0.030) & 14.740(0.030) & $\cdots$ & MITSuME(O) \\
56000.5 & $\cdots$ & $\cdots$ & $\cdots$ & $\cdots$ & 15.282(0.060) & $\cdots$ & $\cdots$ & MITSuME(I) \\
56001.5 & $\cdots$ & $\cdots$ & $\cdots$ & $\cdots$ & $\cdots$ & 14.816(0.034) & $\cdots$ & HOWPol \\
56005.4 & $\cdots$ & $\cdots$ & $\cdots$ & $\cdots$ & 15.616(0.114) & 15.025(0.067) & 15.353(0.243) & HOWPol \\
56006.4 & $\cdots$ & $\cdots$ & $\cdots$ & 16.047(0.139) & 15.550(0.072) & 14.956(0.030) & $\cdots$ & HOWPol \\
56214.7 & $\cdots$ & $\cdots$ & $\cdots$ & $\cdots$ & 20.425(0.304) & 19.625(0.217) & $\cdots$ & HOWPol \\
56223.4 & $\cdots$ & 22.062(0.070) & $\cdots$ & 21.341(0.050) & 21.107(0.096) & 20.070(0.072) & $\cdots$ & FOCAS
   \enddata
\label{op}
\end{deluxetable}{}

  \newpage

\appendix
\section{Overall optical spectral evolutions.}

\begin{figure*}
  \begin{center}
    \begin{tabular}{c}
      \resizebox{120mm}{!}{\includegraphics{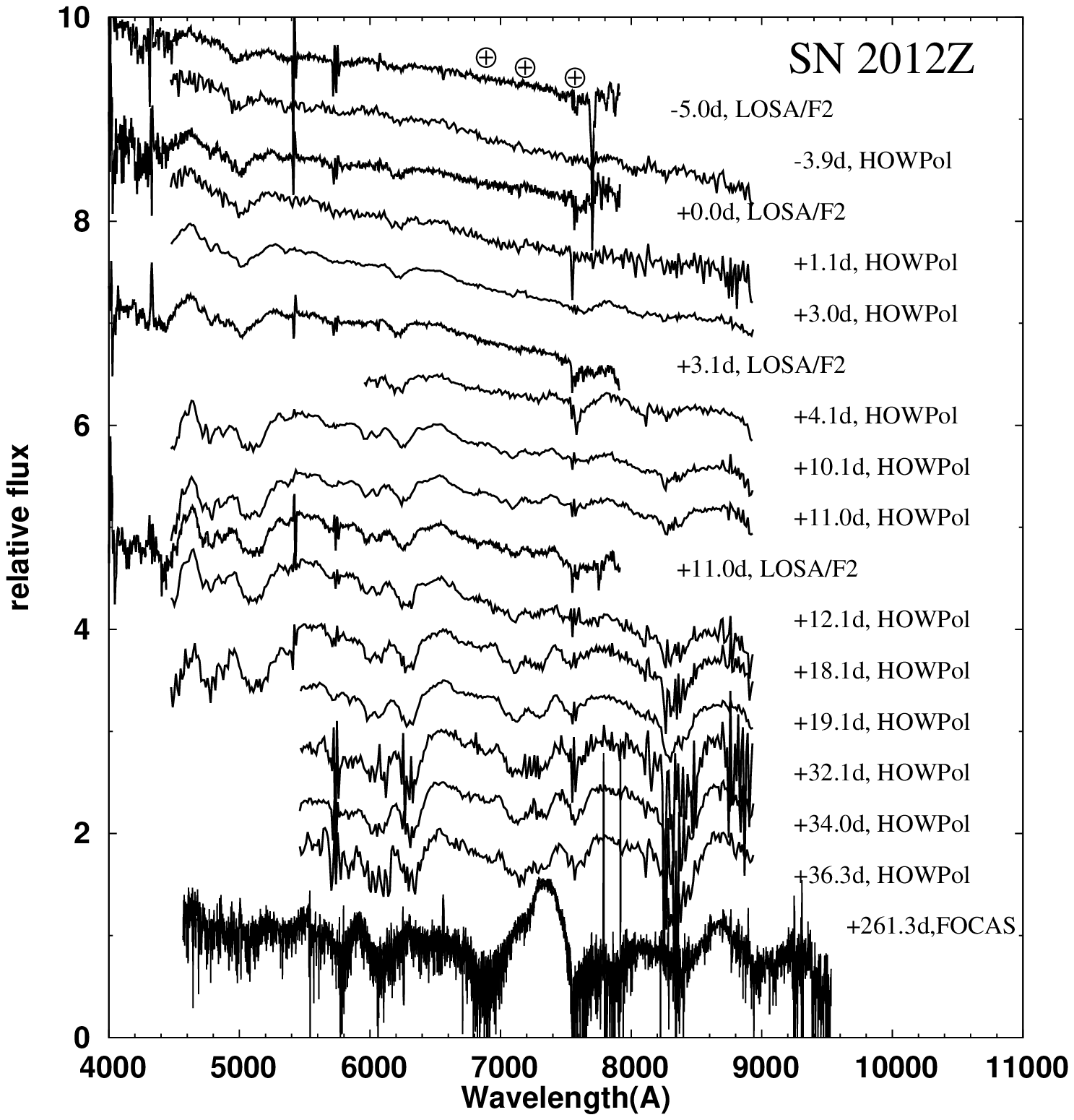}} 
    \end{tabular}
    \caption{Spectral evolutions of SN 2012Z from $t=-5.1$ to 
    $261.3$ d. The absolute flux was artificially shifted. The wavelength 
    was converted to the rest frame using z=0.007. The three symbols denote 
    the wavelength of the atmospheric lines of the Earth. }
    \label{spev}
  \end{center}
\end{figure*}

\end{document}